\documentclass[conference,compsoc]{IEEEtran}
\IEEEoverridecommandlockouts

\usepackage{tikz}
\usepackage{amsmath}
\usepackage{shortcuts}
\usepackage{filecontents}
\usepackage{siunitx}
\usepackage{shortcuts}

\PassOptionsToPackage{table, dvipsnames}{xcolor} 
\usepackage[table]{xcolor}
\usepackage{graphicx}
\usepackage{booktabs,caption}
\usepackage[flushleft]{threeparttable}
\usepackage{tabularx}
\usepackage{makecell}
\usepackage{graphicx} 
\usepackage{pifont}   
\usepackage{wasysym}
\usepackage{arydshln}
\usepackage{tikz}
\usepackage{amsmath}
\usepackage[table]{xcolor}
\usepackage{colortbl} 
\usepackage{siunitx}
\usepackage{makecell}
\usepackage{textcomp}

\usepackage{cite}
\usepackage{amsmath,amssymb,amsfonts}
\usepackage{algorithmic}
\usepackage{graphicx}
\usepackage{textcomp}
\usepackage{xcolor}
\def\BibTeX{{\rm B\kern-.05em{\sc i\kern-.025em b}\kern-.08em
    T\kern-.1667em\lower.7ex\hbox{E}\kern-.125emX}}

\begin{document}
 
\newcommand*\emptycirc[1][1ex]{\tikz\draw (0,0) circle (#1);} 
\newcommand*\quartercirc[1][1ex]{%
  \begin{tikzpicture}
  \draw[fill] (0,0)-- (90:#1) arc (90:180:#1) -- cycle ;
  \draw (0,0) circle (#1);
  \end{tikzpicture}}
\newcommand*\halfcirc[1][1ex]{%
  \begin{tikzpicture}
  \draw[fill] (0,0)-- (90:#1) arc (90:270:#1) -- cycle ;
  \draw (0,0) circle (#1);
  \end{tikzpicture}}
\newcommand*\triquartercirc[1][1ex]{%
  \begin{tikzpicture}
  \draw[fill] (0,0)-- (90:#1) arc (90:360:#1) -- cycle ;
  \draw (0,0) circle (#1);
  \end{tikzpicture}}
\newcommand*\fullcirc[1][1ex]{\tikz\fill (0,0) circle (#1);}

  \date{}

  \title{Devilray: A Reconfigurable Adversarial Baseline for Revealing Blind Spots\\in Fake Base Station Detection}

  \author{\parbox{\linewidth}{\centering Taekkyung Oh$^1$\textsuperscript{*}, Duckwoo Kim$^1$\textsuperscript{*}, Hansung Bae$^2$, Beomseok Oh$^1$, CheolJun Park$^3$, Tyler Tucker$^4$, Nathaniel Bennett$^5$, Sangwook Bae$^6$, Byeongdo Hong$^2$, Patrick Traynor$^4$, Yongdae Kim$^1$\\ \vspace{1mm}
  \small \textit{$^1$KAIST, $^2$The Affiliated Institute of ETRI, $^3$Kyung Hee University, $^4$University of Florida, $^5$Brigham Young University, $^6$Cape}
  }}

  \maketitle
  \def\thefootnote{*}\footnotetext{Both authors contributed equally to this work.}\def\thefootnote{\arabic{footnote}}
  \widowpenalty 0
  \clubpenalty 0
  \begin{abstract}
Fake Base Station (FBS) detection has been a critical focus of cellular security research for over two decades. However, significant financial and regulatory barriers to accessing commercial FBS (C-FBS) devices have limited direct visibility into real-world operations, forcing detection systems to be designed and evaluated around self-built prototypes. 
In this paper, we present \devil, a reconfigurable adversarial baseline to guide effective FBS detection by exploring the realistic adversarial space and identifying blind spots in detection systems -- regions of realistic adversarial behavior excluded by prevailing threat models.
To this end, we establish an empirical anchor through the first academic analysis of a C-FBS, define a fundamental four-phase pipeline and seven operational feature dimensions for adversarial behaviors, and extend them with operational variants derived from 3GPP standards analysis. 
Through combinatorial configurations of operations from commercial practices and specifications, \devil enables the exploration of 2,592 FBS instances, spanning diverse operational possibilities. Using \devil, we evaluate seven currently accessible FBS detectors and uncover coverage gaps across all seven, revealing blind spots rooted in assumption-bound design and evaluation. Our work provides an empirically grounded adversarial baseline for evaluating FBS detectors and guiding future detection design, with \devil and its configuration profiles released to vetted research groups.
\end{abstract}
  \section{Introduction}
 
A Fake Base Station (FBS) impersonates legitimate cellular infrastructure to intercept user connections and conduct privacy and security attacks. Known as Rogue Base Stations, Cell Site Simulators, IMSI-catchers, or Stingrays, these devices are deployed by governments for surveillance~\cite{biddle2016long,gff2020imsi,bristolcable2016imsi,dnc_imsi,privacylegal,uf_imsi_article} and 
exploited by criminals for 
SMS phishing~\cite{sms_blasters_scam,new_risk_smsblaster,chinese_scam,zhang2020lies,smsblaster} and illegal surveillance~\cite{cbc2017imsi,DHS_detection_effort}. Despite advancements in standards, weaknesses in initial connection procedures, the lack of base station authentication, and legacy fallback continue to expose users to tracking, disruption, and message injection.

FBS detection has been studied for over two decades~\cite{meyer2004man,dabrowski2014imsi,ney2017seaglass,samin2024gotta,arnold2024catch}. However, financial and regulatory barriers to accessing commercial FBS (C-FBS) devices~\cite{imsi_catcher_price, css_price_us, eu_imsi_price, jacobs_css_price,stingray_price} have limited direct visibility into real-world adversary operations. Consequently, detectors are commonly designed and evaluated using self-built prototypes with limited signaling and radio configurations. Detectors evaluated under such assumptions may perform well against the modeled adversary while leaving other realistic behaviors unexplored; we refer to these regions as \emph{adversarial blind spots}.

We present \devil, a reconfigurable adversarial baseline for studying such blind spots. We structure \devil into four conceptual layers: (1) the empirical analysis of a single C-FBS as an anchor; (2) a framework comprising a four-phase operational pipeline and seven feature dimensions; (3) specification-derived operational variants extending the framework beyond the observed device; and (4) a reconfigurable implementation with reusable configuration profiles and a dependency checker that validates dependencies and prunes invalid combinations. The single C-FBS does not represent the broader commercial ecosystem, but provides an empirical anchor that we extend using prior work and 3GPP specifications. We focus on LTE because most currently accessible FBS detectors target LTE and its protocol surfaces remain practically exploitable; extending \devil to native 5G-SA requires additional protocol-specific support and is left for future work~(\autoref{s:discussion}).

Within this methodology-bounded space, \devil derives 2,592 FBS configurations. We experimentally validate all 2,592 configurations end-to-end with UEs to confirm their operational feasibility. Rather than treating this space as an exhaustive model of FBS behavior, we use targeted instances from \devil to evaluate seven accessible FBS detectors and identify blind spots arising from assumptions across protocol- and physical-layer behaviors.

We make the following contributions:
\begin{itemize}
\item \textbf{Empirical grounding.} We provide the first academic analysis of a C-FBS, using it as an empirical anchor to identify gaps between academic assumptions and observed real-world FBS operations.

\item \textbf{Adversarial framework and implementation.} We develop a four-phase framework with seven feature dimensions, extend it with specification-derived variants, and realize it as the reconfigurable \devil implementation.

\item \textbf{Detector evaluation and blind-spot analysis.} We evaluate seven accessible FBS detectors using targeted \devil configurations and characterize adversarial blind spots in their coverage. Demo videos of representative attack workflows are available at our website~\cite{devilray_demo}.

\item \textbf{Artifact release and practical adoption.} We release \devil and reusable configuration profiles to vetted researchers as a shared baseline for reproducible FBS detection research. \devil has been shared with the Electronic Frontier Foundation (EFF) to evaluate and improve Rayhunter~\cite{rayhunter} at their request, and we are conducting experiments with \devil on a carrier testbed in collaboration with SK Telecom, a major Korean carrier.
\end{itemize}

\nibf{Ethics.}
We obtained temporary access to a C-FBS (\$50K) from a third party. All active experiments requiring signal generation involving the C-FBS, \devil, and target phones were conducted only in shielded, access-controlled laboratories enclosed in conductive mesh to prevent signal leakage. A cellular repeater relayed legitimate 2G--LTE signals to enable normal phone operation during experiments. Passive measurements were conducted without shielding. Due to dual-use concerns surrounding \devil's reusable, turnkey attack configurations, the codebase, dependency checker, configuration profiles, and trace datasets are archived under controlled access and accessible only to vetted researchers.
  \section{Background}
\label{s:background}
\nibf{LTE architecture.}
An LTE network comprises three main components: the User Equipment (UE), Radio Access Network (RAN), and Core Network. UEs connect to base stations (eNBs) via wireless channels managed by the RAN, while the Core Network handles authentication, session management, and mobility control. UEs interact with the network via Radio Resource Control (RRC) with the eNB and Non-Access Stratum (NAS) with the core network.

\begin{figure}[t]
    \centering
    \includegraphics[width=.8\columnwidth]{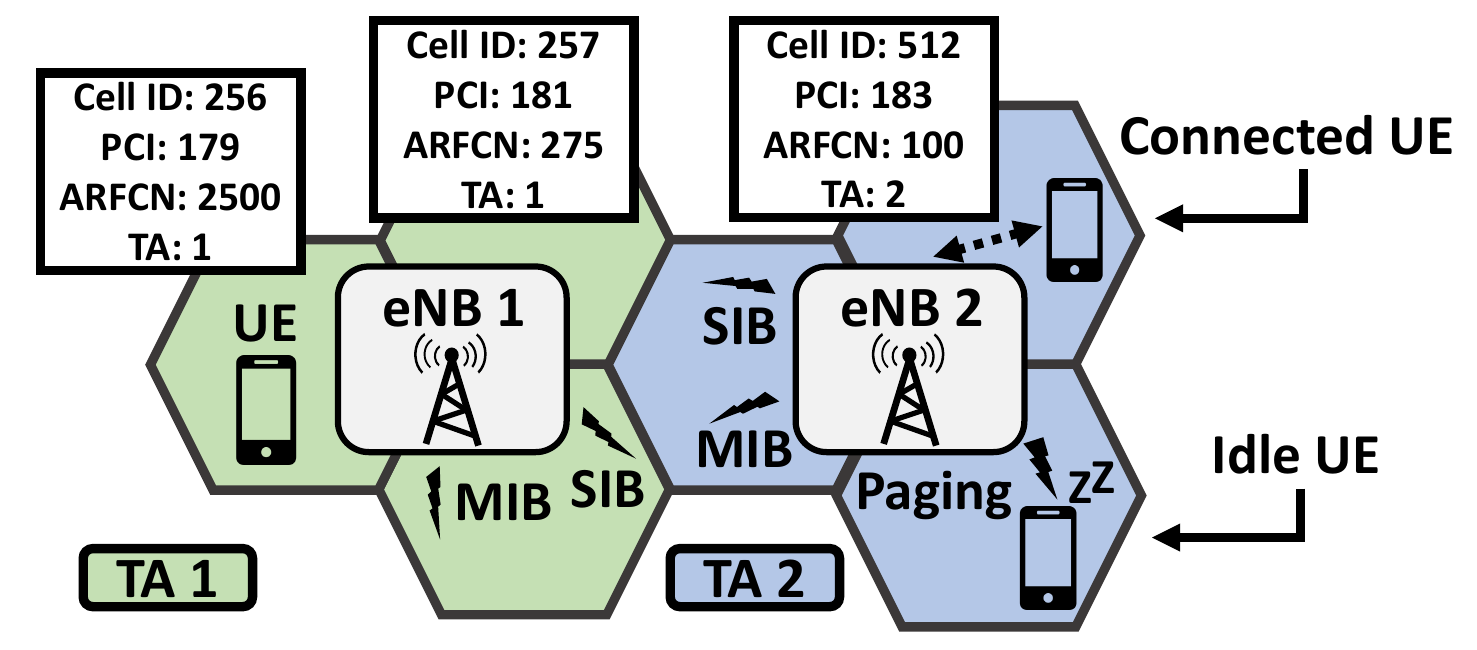}
    \caption{LTE architecture and broadcast messages}
    \label{fig:lte_overview}
\end{figure}

\nibf{Network identifier.}
LTE cells broadcast identifiers for network recognition and mobility management~(\autoref{fig:lte_overview}). Key identifiers include a Public Land Mobile Network (PLMN) for operator identification and a Tracking Area Code (TAC) defining logical regions for paging. Each cell is distinguished by a network-unique Cell Identity (Cell ID) and a Physical Cell Identity (PCI) for synchronization and initial access. Cells operate on frequency channels defined by Absolute Radio Frequency Channel Numbers (ARFCN), from which UEs derive the PCI through synchronization signals.

\nibf{Broadcast message.}
eNBs broadcast Master Information Blocks (MIB) for frame timing and bandwidth, and System Information Blocks (SIB1 to SIB24). SIB1 carries the PLMN ID, TAC, cell ID, and scheduling for other SIBs, while SIB3 (intra-frequency) and SIB5 (inter-frequency) dictate cell reselection priorities. Additionally, the network broadcasts paging messages to notify idle UEs of incoming services, prompting connection establishment.

\nibf{Mobility management.}
LTE handles UE mobility based on its operational state. In the idle state, UEs monitor broadcast messages and measure neighboring signal quality to execute cell reselection, camping on the optimal cell based on priority criteria. In the connected state, UEs utilize handover to maintain seamless connectivity. Handover is network-initiated based on periodic or signal-triggered UE measurement reports; if a target cell exhibits significantly better quality, the network transfers the UE while forwarding session and radio configurations.

  \section{Motivation}
\label{s:motivation}

To provide the motivation for our work, we address four fundamental research questions throughout this paper.
While this section focuses on uncovering the current gaps (RQ1 \& RQ2), the subsequent sections investigate the resulting threat modeling requirements (RQ3) and blind spots (RQ4):
\begin{itemize}
    \item \textbf{RQ1. Operational reality}: How does real-world FBS actually operate in practice?
    \item \textbf{RQ2. Academic assumptions}: Which academic assumptions are violated by commercial FBS operations? 
    \item \textbf{RQ3. Threat modeling requirements}: What requirements must adversarial models satisfy to enable effective and realistic detection research?  
    \item \textbf{RQ4. Security risks:} How do these gaps translate into security risks when encountering realistic FBS threats?
\end{itemize}

\subsection{FBS in Practice (RQ1)}
\label{ss:c-fbs}
We analyze a C-FBS sold to law enforcement and intelligence agencies for ~\$50K (product name undisclosed under confidentiality agreement) to understand real-world threat capabilities. Using UE-side diagnostic logs via SCAT~\cite{hong2018peeking} and C-FBS internal logs, we capture its full-chain behavior. 

\nibf{Integrated hardware architecture.}
\label{ss:cfbs_hw}
The C-FBS features a Software Defined Radio (SDR) with an onboard processor for mobile, remote operation, integrating a dual-channel Radio Frequency (RF) transceiver, Wi-Fi module, and Android control app. Supporting 2G/3G/LTE across 400 MHz–6 GHz, each channel can configure an independent base station at different frequencies. This enables both multi-RAT (Radio Access Technology) deployment and concurrent dual-channel operation within the same RAT. Despite the integrated architecture, its bandwidth remains fixed at 5 MHz due to hardware constraints. It includes a GPS module for clock synchronization and external storage for logs.

\nibf{Operational logic.}
\label{ss:cfbs_operation}
The C-FBS provides a three-phase operational workflow designed for non-experts, bundling protocol steps into sequential, operator-facing phases:

\niit{i) Reconnaissance:} To identify target networks, it profiles radio environments, scanning nearby cells for PLMN, RAT, PCI, cell ID, ARFCN, TAC, and signal strength. This data is stored to guide subsequent configuration, supplemented by preset network configurations for various countries.

\niit{ii) Automated configuration:} Using scanned data, it targets regional bands and automatically initializes broadcast identifiers (\eg, PCI, TAC, cell ID) with random values. This baseline defaults to IMSI-catching as the primary vector, before enabling location tracking, downgrading, or SMS injection using captured IMSIs.

\niit{iii) Attack execution:}
The C-FBS activates RF output, broadcasting MIB and SIB (type 1-5) messages to lure victims. Upon connection, it executes IMSI-catching via an \tts{Identity Request}, logs captured IMSIs, and disconnects via a \textit{NAS Reject} message. It then supports location tracking (triggering uplink traffic via paging and \tts{Identity Request} to measure signal strength and propagation delay for distance estimation) or downgrading and SMS injection by forcing the UE to its concurrent 2G cell via RRC redirection and exploiting unencrypted links.

\nibf{Extended adversarial capabilities.}
\label{ss:cfbs_advanced}
Beyond well-known operational logic, the C-FBS supports extensive control over attack parameters, enabling combinatorial configurations for diverse operational strategies.

\niit{i) Flexible configuration:} Operators can manually configure target bands and broadcast identifiers to clone legitimate cells for stealth (environment adaptation). It also supports protocol message tailoring by selecting \textit{NAS Reject} causes (\eg, \#13 \textit{Roaming not allowed}, \#22 \textit{Congestion}, or \#42 \textit{Severe network failure}~\cite{3gpp_24301}) to control the victim's state --  blacklisting TACs, inducing DoS~\cite{liu2024dark}, or forcing redirection~\cite{karakoc2023never} to its concurrent 2G/3G cell for multi-step attacks.

\niit{ii) Adaptive temporal and frequency hopping:} It enables automatic migration across different frequencies and RATs with configurable dwell times. This sweeps multiple bands to maximize targets while confusing static detection due to its transient nature and distribution of signals.

\niit{iii) Multi-RAT orchestration:} The dual-channel architecture supports arbitrary network combinations concurrently, allowing it to mirror legitimate topologies and stealthily expand the attack surface. Operators can deploy heterogeneous setups (LTE-to-2G for conventional downgrades; LTE-to-3G to target modern UEs supporting 2G disable options~\cite{eff_null,google_2g}) or homogeneous instances (\eg, dual LTE cells).

\niit{iv) Dynamic transmission management:} Adjustable power and Tx attenuation tune the effective range (within 2–10 m) to balance stealth and coverage efficiency, while supporting external amplifiers to boost range. Combined with flexible cell configuration capabilities, this facilitates covert UE luring via handover and reselection mechanisms~(\autoref{ss:evasion}).
  \subsection{Limitations in Prior Threat Models (RQ2)}
\label{ss:limitations}

\begin{table}[t]
\centering
\caption{Gaps in academic FBS models, compared to C-FBS.}
\label{tab:threat_model_comparison}
\setlength{\tabcolsep}{2pt}
\resizebox{\linewidth}{!}{
\begin{tabular}{lccccccc}
\toprule
 & \multicolumn{5}{c}{\textbf{Adversarial capabilities of C-FBS~(\autoref{ss:cfbs_advanced})}} & \textbf{Modeling} & \textbf{Reference} \\ 
\cmidrule(lr){2-6} \cmidrule(lr){7-7} \cmidrule(lr){8-8}
 & \multicolumn{2}{c}{\textbf{Flexible config.}} & & & & & \\
\cmidrule(lr){2-3}
& \textbf{Cell-clone} 
& \textbf{Msg-varying} 
& \multirow{-2.5}{*}{\textbf{\makecell{Freq.\\hopping}}} 
& \multirow{-2.5}{*}{\textbf{\makecell{Multi-RAT\\orchestr.}}} 
& \multirow{-2.5}{*}{\textbf{\makecell{Dynamic\\transm.}}} 
& \multirow{-2.5}{*}{\textbf{\makecell{Full-chain\\view$^\ast$}}}
& \multirow{-2.5}{*}{\textbf{\makecell{Artifact\\avail.}}} \\
\midrule
Ali \etal~\cite{ali2019enabling}                    & \emptycirc & \emptycirc & \emptycirc & \emptycirc & \emptycirc & \emptycirc & \emptycirc \\ 
Apple's patent~\cite{apple_patent}                 & \emptycirc & \emptycirc & \emptycirc & \fullcirc  & \emptycirc & \emptycirc & \emptycirc \\ 
Bitsikas \etal~\cite{bitsikas2021don}              & \fullcirc  & \emptycirc & \emptycirc & \fullcirc  & \fullcirc  & \emptycirc & \emptycirc \\ 
CellGuard~\cite{arnold2024catch}                   & \emptycirc & \emptycirc & \emptycirc & \emptycirc & \emptycirc & \emptycirc & \emptycirc \\
Crocodile Hunter~\cite{crocodilehunter}            & \emptycirc & \emptycirc & \emptycirc & \emptycirc & \emptycirc & \emptycirc & \emptycirc \\ 
\addlinespace[1.5pt]
FBS-Radar~\cite{li2017fbs}                         & \fullcirc  & \emptycirc & \emptycirc & \fullcirc  & \emptycirc & \emptycirc & \emptycirc \\ 
FBSDetector~\cite{samin2024gotta}                  & \fullcirc  & \fullcirc  & \emptycirc & \fullcirc  & \fullcirc  & \emptycirc & \halfcirc$^\dagger$ \\
FBSleuth~\cite{zhuang2018fbsleuth}                 & \emptycirc & \emptycirc & \fullcirc  & \emptycirc & \emptycirc & \emptycirc & \emptycirc \\
FlashCatch~\cite{paci2025flashcatch}               & \emptycirc & \fullcirc  & \emptycirc & \emptycirc & \fullcirc  & \emptycirc & \emptycirc \\
GSMK Overwatch~\cite{projectoverwatch}             & \fullcirc  & \emptycirc & \emptycirc & \fullcirc  & \emptycirc & \emptycirc & \emptycirc \\ 
\addlinespace[1.5pt]
Heijligenberg \etal~\cite{heijligenberg2024attacks}& \emptycirc & \fullcirc  & \emptycirc & \emptycirc & \fullcirc  & \emptycirc & \emptycirc \\ 
Huang \etal~\cite{huang2018identifying}            & \emptycirc & \emptycirc & \emptycirc & \emptycirc & \emptycirc & \emptycirc & \emptycirc \\ 
IMSI-catcher-catcher~\cite{dabrowski2014imsi}      & \emptycirc & \emptycirc & \emptycirc & \fullcirc  & \emptycirc & \emptycirc & \emptycirc \\ 
LeopardSeal~\cite{peeters2023leopardseal}          & \emptycirc & \emptycirc & \emptycirc & \fullcirc  & \emptycirc & \emptycirc & \emptycirc \\ 
Marlin~\cite{tucker2025detecting}                  & \emptycirc & \fullcirc  & \emptycirc & \fullcirc  & \emptycirc & \emptycirc & \halfcirc$^\dagger$ \\
\addlinespace[1.5pt]
Murat~\cite{nakarmi2021murat}                      & \emptycirc & \emptycirc & \emptycirc & \fullcirc  & \emptycirc & \emptycirc & \emptycirc \\ 
Park~\cite{park2023we}                             & \fullcirc  & \fullcirc  & \emptycirc & \fullcirc  & \fullcirc  & \emptycirc & \emptycirc \\ 
Phoenix~\cite{echeverria2021phoenix}               & \fullcirc  & \fullcirc  & \emptycirc & \emptycirc & \emptycirc & \emptycirc & \halfcirc$^\dagger$ \\
Rayhunter~\cite{rayhunter}                         & \emptycirc & \emptycirc & \emptycirc & \fullcirc  & \emptycirc & \emptycirc & \emptycirc \\ 
Seaglass~\cite{ney2017seaglass}                    & \fullcirc  & \emptycirc & \emptycirc & \fullcirc  & \emptycirc & \emptycirc & \emptycirc \\ 
\addlinespace[1.5pt]
Shaik \etal~\cite{shaik2018impact}                 & \fullcirc  & \emptycirc & \emptycirc & \emptycirc & \fullcirc  & \emptycirc & \emptycirc \\
Steig \etal~\cite{steig2016network}                & \emptycirc & \emptycirc & \emptycirc & \emptycirc & \emptycirc & \emptycirc & \emptycirc \\ 
\bottomrule
\end{tabular}
}
\begin{tablenotes}
    \scriptsize
    \item[*] $\ast$ Comprehensive understanding from physical-layer features dictated by hardware to complex protocol behaviors.
    \item[*] $\dagger$ Provide traffic gathered from their own FBS implementation or field measurements.
\end{tablenotes}
\end{table}

We identify limitations in academic adversary modeling by comparing prior models against our C-FBS analysis to reveal practical gaps between academic assumptions and real-world capabilities~(\autoref{tab:threat_model_comparison}). This section focuses on characterizing threat modeling rather than detection mechanisms.

First, academic models 
employ self-built prototypes with simplified behaviors: static adversaries with predictable patterns such as distinctive cell identifiers, constant high power, and static frequency residency. 
Our C-FBS analysis reveals its operational capabilities are absent from most models, including environment adaptation, frequency hopping, and dynamic transmission.
This limited scope also prevents the exploration of combinatorial configurations across the diverse operations observed in C-FBS, restricting prior models to case-specific threat designs.
While some studies~\cite{samin2024gotta,park2023we,echeverria2021phoenix} attempted to incorporate operational diversity with flexible configuration, \autoref{tab:threat_model_comparison} reveals most still provide only partial coverage across operational dimensions: implementing subsets of flexible configuration, frequency hopping, multi-RAT, or dynamic transmission without cohesive integration enabling operational diversity.

Additionally, academic models typically represent isolated aspects of FBS operations rather than integrated attack chains. \autoref{tab:threat_model_comparison} shows existing models lack full-chain coverage spanning (hardware-originated) physical-layer features, protocol-layer behaviors, and multi-phase workflows (reconnaissance to attack execution). Notably, our C-FBS analysis reveals these elements operate in concert: hardware constraints influence protocol behaviors (\eg, bandwidth information and multi-RAT RF-channels), attack phases depend on prior configurations, and operational success requires end-to-end coordination. Fragmented models miss these cross-component dependencies observed in C-FBS.

\begin{figure*}[t]
    \centering
    \includegraphics[width=.95\linewidth]{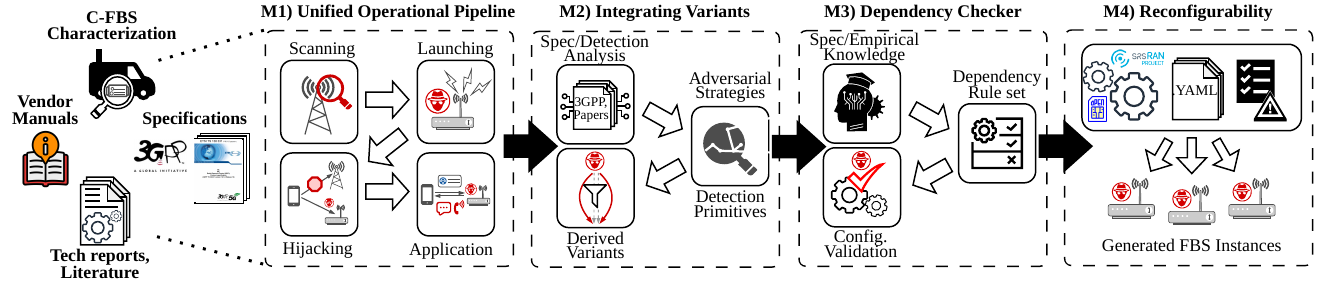}
        \caption{The \devil methodology (M1--M4).}
    \label{fig:devilray_methodology}
\end{figure*}

\nibf{Implications.} These gaps can create adversarial blind spots in detection. Detection research relying on simplified threat models 
optimizes against narrow adversarial assumptions, potentially overlooking operational capabilities available in commercial FBS implementations. Additionally, most academic models remain unavailable as artifacts~(\autoref{tab:threat_model_comparison}), hindering research continuity. Without accessible reference implementations, each study develops isolated prototypes, preventing cumulative progress and objective benchmarking.
  \section{Approach: \deviltitle}
\label{s:approach}
To address modeling gaps in~\autoref{s:motivation}, 
we establish threat modeling requirements for FBS research and introduce \devil that satisfies them with structured methodologies. 

\subsection{Threat Modeling Requirements (RQ3)}
\label{ss:requirements}
Based on gaps identified in~\autoref{s:motivation}, we derive four requirements for adversarial modeling. They address both detection evaluation and representation of commercial threats, including high-end systems reportedly priced up to \$1M+~\cite{css_jacob_brochure}.

\nibf{R1: Establishing unified baseline with empirical ground.} 
While prior modeling efforts have largely relied on simplified and fragmented prototypes, a unified baseline spanning end-to-end FBS operations is required to model an extensive threat space and provide a coherent view.
Moreover, to provide a concrete foundation rooted in actual threats, this baseline should be empirically grounded in real-world operational capabilities, such as those revealed in our C-FBS analysis~(\autoref{ss:c-fbs}). 
While scoped to a single device and not representative of all products, it provides an empirical anchor in real-world behaviors, not hypothetical modeling.

\nibf{R2: Capturing operational realism.}
Analyzing all commercial FBS devices is infeasible. Therefore, threat modeling requires 
exploring the adversarial configuration space -- operations adversaries can perform within protocol constraints defined in 3GPP standards -- to capture broader operational realism beyond the single-device empirical observations via combinatorial configurations.
This includes uncovering standard-compliant alternatives (\eg, alternative IMSI-catching like \textit{NAS Reject} instead of \tts{Identity Request}~\cite{tucker2025detecting}) and strategic adaptation into operational~contexts (\eg, fine-tuning cell parameters or reducing attack volumes) rather than narrow assumptions and fixed behaviors.

\nibf{R3: Ensuring adversarial feasibility.}
Indiscriminate exploration of the configuration space, including combinations of diverse alternatives and strategies, can introduce invalid adversarial operations -- such as ineffective UE luring and violations of 3GPP standards. Thus, the threat model should account for protocol semantics, hardware constraints, and operational dependencies to prune invalid configurations. Otherwise, it risks producing inoperable setups, undermining the model's practical realism and evaluation utility.

\nibf{R4: Enabling reproducible instantiation.}
To support detection research effectively, the model should not only describe diverse adversaries but also instantiate them as reproducible threat profiles; given that the commercial FBS ecosystem is heterogeneous, spanning \$10K portable units to \$1M+ integrated systems~\cite{css_jacob_brochure} with distinct radio resources and feature sets. 
Therefore, it requires a reconfigurable system capable of instantiating a wide spectrum of adversarial configurations in an end-to-end manner.

\subsection{Methodology of \devil}
\label{ss:methodology}
We build \devil with the following methodologies to address these requirements, as illustrated in~\autoref{fig:devilray_methodology}:

\nibf{M1: Empirically grounded operational pipeline.} We adopt a unified four-phase operational pipeline consisting of scanning, launching, hijacking, and application, derived from analysis of literature and our C-FBS study. The C-FBS workflow~(\autoref{ss:cfbs_operation}) is operator-facing and coarse-grained; \devil refines execution into launching, hijacking, and application to separate operational stages and enable phase-specific configuration. Across this pipeline, we also establish seven feature dimensions to characterize fundamental adversarial operations. This framework captures an end-to-end view spanning physical- and protocol-layer behaviors.

\nibf{M2: Specification-driven variation derivation.} 
To model diverse, realistic adversarial strategies beyond single-device anchoring, \devil integrates operational variations and adaptations across the seven feature dimensions. These are derived from the analysis of detection primitives (standardized artifacts monitored by detectors) and 3GPP spec-based adversarial strategies, using the former as an indexing lens to scope relevant 3GPP clauses across dimensions rather than introduce detector-tailored variants. This captures standard-compliant alternatives that adversaries can exploit in practice, avoiding ad-hoc or hypothetical assumptions while enabling principled expansion of the adversarial design space.

\nibf{M3: Configuration validation.}
\devil incorporates a dependency checker that filters out logical conflicts yielding invalid configurations by validating protocol semantics, hardware constraints, and operational dependencies based on empirical and domain knowledge. The checker encodes these conflicts into an initial baseline of 21 rules, semantically identifying and automatically resolving conflicts to ensure the logical exploration of the adversarial space.
For a tractable yet principled scope, this validation is bounded by 3GPP specifications and our seven feature dimensions, while the rule set remains operator-configurable to accommodate target deployment constraints.

\nibf{M4: Reconfigurable implementation.}
\devil is built as a reconfigurable system that supports modular instantiation across the adversarial configuration space. This enables the exploration of 2,592 distinct FBS instances spanning diverse, realistic adversarial behaviors. It supports storing and reusing these instances via YAML-formatted configuration profiles, which will be released alongside the model for reproducible research.
  \section{Design of \deviltitle}
\label{s:devil_design}
\begin{figure*}[t]
    \centering
    \includegraphics[width=.95\textwidth]{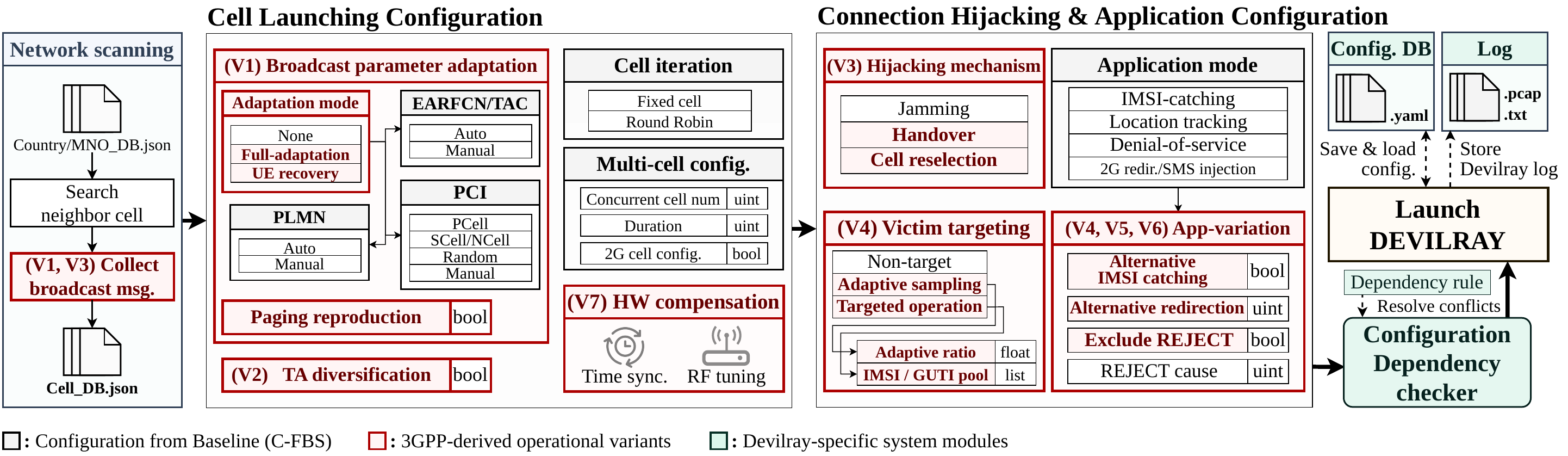}
        \caption{\devil's modular and reconfigurable architecture, showing four operational phases and seven variant dimensions.
        }
    \label{fig:devilray_config}
\end{figure*}
We design \devil based on our methodology. We first establish a unified four-phase operational pipeline~(\autoref{ss:baseline}) extending C-FBS capabilities and characterizing core operational feature dimensions. Building on this, we derive and integrate operational variants~(\autoref{ss:evasion}) through a specification-driven expansion of feature dimensions based on detection domain analysis and 3GPP standards.
We then introduce a dependency checker~(\autoref{ss:dep_checker}) to filter invalid configurations and enable logical exploration of configuration-space diversity. Finally, we present a reconfigurable implementation~(\autoref{ss:implementation}) enabling instantiation of 2,592 configurations spanning realistic adversarial capabilities. The full spectrum of \devil, including its modular, reconfigurable architecture and variants, is illustrated in~\autoref{fig:devilray_config}.

\subsection{Unified Four-Phase Operational Pipeline}
\label{ss:baseline}
We establish a unified operational pipeline of FBS that consists of four phases -- network scanning, cell launching, connection hijacking, and application -- and design fundamental adversarial behaviors of \devil for each phase.
This pipeline consolidates fragmented and inconsistent views from prior literature into a cohesive framework where each phase produces specific outputs that constrain subsequent operations, guaranteeing operational consistency.

To construct this pipeline, we synthesize knowledge from a broad range of sources:
literature on cellular surveillance~\cite{eff_surveillance,nasser2019gotta,css_jacob,privacylegal,eu_imsi_price,jacobs_css_price},
academic studies~\cite{samin2024gotta,ney2017seaglass,dabrowski2014imsi,crocodilehunter,projectoverwatch,li2017fbs,nakarmi2021murat,karaccay2021network,heijligenberg2024attacks,nakarmi2022applying,park2023we,arnold2024catch,huang2018identifying,tucker2025detecting,rayhunter,echeverria2021phoenix,steig2016network,ali2019enabling,zhuang2018fbsleuth,bitsikas2021don,yang2019hiding,shaik2018impact,paci2025flashcatch,zhang2020lies},
3GPP specifications~\cite{3gpp_36304,3gpp_36331,3gpp_33809}, technical reports~\cite{dnc_imsi,bristolcable2016imsi,gff2020imsi,DHS_detection_effort,uf_imsi_article}, publicly available vendor manuals~\cite{biddle2016long,vendor1,vendor2,vendor3,vendor4,vendor5}, and C-FBS~(\autoref{ss:c-fbs}).

\nibf{Network scanning.}
\label{ss:network_scanning}
\devil begins with reconnaissance to identify target networks with which nearby UEs are communicating and gather configuration details from legitimate cells, necessary for subsequent cell setup and impersonation of those cells. Collected information is stored in a JSON-based database for later use.

In this phase, \devil 
scans available frequency bands to passively collect broadcast information, 
including MIB, all SIBs (type 1-24), and paging patterns from legitimate cells. Beyond the basic metadata (\eg, PLMN, PCI, signal strength) captured by C-FBS, \devil harvests fine-grained parameters, including random access configurations, scheduling settings, and cell reselection properties. 

\nibf{Cell launching.}
\devil offers a configuration space for fine-grained cell parameters, broadcast signals, and RF operations to emulate benign base stations. This phase involves strategic decisions on cell identifiers, power levels, and air-interface settings to make nearby UEs recognize \devil as a viable candidate.

\niit{i) Configuration of cell parameters:} 
\devil initializes exhaustive broadcast identifiers and radio configurations, including PLMN, TAC, PCI, bandwidth, ARFCNs, paging content, and reselection priorities. \devil supports the complete definition of these settings by utilizing harvested data during the scanning phase to populate cellular configurations.
This capability facilitates granular control over comprehensive broadcast and unencrypted channel information, surpassing that of C-FBS to reflect network environments.

\niit{ii) Cell-iteration:}
\devil supports two operational methods: fixed or round-robin. Fixed mode maintains static cell parameters (PLMNs, ARFCNs, PCIs) throughout a dwelling duration. Round-robin mode rotates through different cell parameters at each interval, cycling across multiple PLMN-ARFCN–PCI combinations with frequency hopping to target different bands while distributing signal patterns.

\niit{iii) Multi-cell:} 
\devil can utilize its multi-channel capability with multiple SDRs to support concurrent multi-cell operations across homogeneous or heterogeneous RATs. 

\nibf{Connection hijacking.}
\label{ss:hijacking}
\devil provides configurable adversarial strategies in this phase by supporting three hijacking mechanisms to lure nearby UEs: jamming, handover, and cell reselection.
Here we address jamming as the baseline hijacking strategy, which operates with significant transmission power to lure UEs.
Beyond jamming, other two mechanisms are extended as operational variants in \autoref{ss:evasion}, which are grounded in specification and empirical study (see~\autoref{appendix:eval_efficacy} for details). The detailed message flows for each hijacking methodology are illustrated in \autoref{fig:connection_hij_flow} (Appendix).

The basic strategy of Jamming\footnote{Unlike conventional jamming as a means to force UEs to migrate to different frequency bands through interference, \devil's jamming refers to overwhelming legitimate signals on the same frequency with a power advantage, exploiting cell selection logic to directly hijack UEs.} that \devil provides is to force UEs to disconnect from legitimate cells and redirect them by overwhelming legitimate signals with higher-power transmissions. 
This exploits that UEs inherently prefer the strongest available cell~\cite{eff_surveillance}.

The specific effect of hijacking depends on the UE’s state. 
For connected UEs, jamming triggers a Radio Link Failure (RLF) and the transmission of an \tts{RRC Reestablishment Request} to \devil, which fails because \devil lacks the UE context, ultimately redirecting them to \devil through a new connection. 
For idle mode UEs, which camp on the serving cell by listening to broadcast messages, jamming disrupts the reception of these messages, causing them to eventually camp on \devil's broadcasts. Although effective against both UE states, jamming requires substantial transmission power ($>$ 20\,dB than benign cells), limiting its attack range. 

\nibf{Application.}
\label{ss:application}
\devil provides four primary attacks:

\niit{i) IMSI-catching:}
The fundamental IMSI-catching functionality collects the UE's IMSI by sending an \tts{Identity Request} message and then promptly disconnects the UE via a \textit{NAS Reject} message. Beyond this, it also supports further variants, which will be detailed in~\autoref{ss:evasion}.

\niit{ii) Location tracking:}
Using IMSIs, \devil can track UEs' locations by periodically issuing IMSI paging and \tts{Identity Request} messages to trigger UE responses and measure the response signal strength. This design, inspired by C-FBS, allows for both coarse-grained presence checks and fine-grained localization with a directional antenna.

\niit{iii) DoS:}
\devil can disrupt UEs' network access by sending a crafted \textit{NAS Reject} with cause \#22~\cite{3gpp_24301}, which force UEs to suspend network search for extended periods (up to 30 minutes)~\cite{liu2024dark}. By repeatedly injecting such messages, it can maintain persistent service unavailability.

\niit{iv) 2G redirection and SMS injection:}
\devil can redirect hijacked UEs from LTE to legacy 2G bands via unprotected RRC redirection procedures (\tts{redirected CarrierInfo} in \tts{RRC Connection Release}) or inter-RAT cell reselection (SIB7). This enables \devil to perform SMS injection, known as the SMS blaster~\cite{smsblaster}, which exploits the weak authentication and encryption of 2G networks. By operating both LTE and 2G channels simultaneously, \devil hijacks the UE on LTE, redirects it to 2G, and performs SMS injection. Also, \devil includes an alternative redirection vector (\autoref{ss:evasion}).

\begin{table}[t]
\centering
\caption{FBS operational features and corresponding \devil's variants based on related standards.}
\label{tab:adversarial_mapping_vertical}
\setlength{\tabcolsep}{4pt}
\resizebox{.9\linewidth}{!}{%
\begin{tabular}{lccc}
\toprule
\textbf{Feature dimension} & \textbf{Phase} & \textbf{Related Standards} & \textbf{Variants} \\ 
\midrule
Cell parameters            & P1/P2          & TS 24.301~\cite{3gpp_24301}, TS 36.331~\cite{3gpp_36331} & \textbf{V1} \\
Restricted coverage$^\ast$ & P2/P3          & TS 36.321~\cite{3gpp_36321}                               & \textbf{V2} \\
Signal power               & P1/P3          & TS 36.304~\cite{3gpp_36304}, TS 36.331~\cite{3gpp_36331} & \textbf{V3} \\
Identity signaling         & P4             & TS 24.301~\cite{3gpp_24301}                               & \textbf{V4} \\
Legacy redirection         & P4             & TS 36.331~\cite{3gpp_36331}                               & \textbf{V5} \\
Protocol sequence          & P4             & TS 24.301~\cite{3gpp_24301}, TS 36.331~\cite{3gpp_36331} & \textbf{V6} \\
HW dependency              & P2             & TS 36.401~\cite{3gpp_36401}, TS 38.401~\cite{3gpp_38401} & \textbf{V7} \\ 
\bottomrule
\end{tabular}%
}\par
\parbox{.9\linewidth}{\raggedright \scriptsize
$\ast$It defines a logical cell boundary enforced via protocol mechanisms such as Timing Advance (TA) under inherent hardware and power constraints.}
\vspace{-5mm}
\end{table}

\nibf{Operational feature dimensions.}
Based on the four-phase pipeline, we define seven feature dimensions to characterize fundamental adversarial operations distinguishable from normal network behaviors, as summarized in~\autoref{tab:adversarial_mapping_vertical}. These dimensions capture the essential operational aspects of an FBS across the cellular protocol stack and the physical layer. Each dimension corresponds to a functional axis mapped to specific phases within the operational pipeline. By decomposing adversarial behavior into these modular dimensions, \devil establishes a structured baseline that scopes the operational space, from which variants are derived and explored, as detailed in~\autoref{ss:evasion}.

\subsection{Integrating Operational Variants}
\label{ss:evasion}

Building on \devil's operational pipeline and feature dimensions, we derive and integrate operational variants that span realistic adversarial capabilities beyond the single empirical observation. 
These variants represent specification-compliant operational diversity. 
\providecommand{\pcirc}{\raisebox{0.15ex}{\scalebox{0.55}{\fullcirc}}}

\begin{table}[t]
\centering
\caption{Detection primitives across FBS detection systems and operational coverage of \devil's variants.}
\label{tab:fbs_detection_primitives}
\setlength{\tabcolsep}{3.8pt}
\resizebox{\linewidth}{!}{%
\scriptsize
\begin{tabular}{l | cccccccccc}
\toprule
& 
\rotatebox{90}{\makecell[l]{Signal strength}} & 
\rotatebox{90}{\makecell[l]{RRC failure}} & 
\rotatebox{90}{\makecell[l]{IMSI-exposing}} & 
\rotatebox{90}{\makecell[l]{Reject message}} & 
\rotatebox{90}{\makecell[l]{Cell info.}} & 
\rotatebox{90}{\makecell[l]{Redirection}} & 
\rotatebox{90}{\makecell[l]{Null cipher}} & 
\rotatebox{90}{\makecell[l]{TA command}} & 
\rotatebox{90}{\makecell[l]{RF char.}} & 
\rotatebox{90}{\makecell[l]{Timing err.}} \\
\midrule
Ali \etal~\cite{ali2019enabling}                  & & & & & & & & & \pcirc & \\ \hline
Apple patent~\cite{apple_patent}                  & \pcirc & & & & \pcirc & \pcirc & & \pcirc & \pcirc & \pcirc \\ \hline
CellGuard~\cite{arnold2024catch}                  & \pcirc & & & \pcirc & \pcirc & & & & & \\ \hline
Crocodile Hunter~\cite{crocodilehunter}           & \pcirc & & & & \pcirc & & & & & \\ \hline
EAGLE Security~\cite{eaglesecurity}               & \pcirc & & & & \pcirc & & & & & \\ \hline
FBSDetector~\cite{samin2024gotta}                 & \pcirc & \pcirc & \pcirc & \pcirc & \pcirc & & \pcirc & & & \\ \hline
FBSleuth~\cite{zhuang2018fbsleuth}                & & & & & & & & & \pcirc & \\ \hline
FBS-Radar~\cite{li2017fbs}                        & \pcirc & & & & \pcirc & & & & & \\ \hline
GSMK Overwatch~\cite{projectoverwatch}            & & & & & \pcirc & \pcirc & \pcirc & & & \\ \hline
Heijligenberg \etal~\cite{heijligenberg2024attacks}& \pcirc & & & & \pcirc & & & & & \\ \hline
Huang \etal~\cite{huang2018identifying}           & \pcirc & & & & & & & & & \\ \hline
IMSI-Catcher-Catcher~\cite{dabrowski2014imsi}     & \pcirc & & & & \pcirc & & \pcirc & & & \\ \hline
Kara{\c{c}}ay \etal~\cite{karaccay2021network}    & \pcirc & & & & & & & & & \\ \hline
LeopardSeal~\cite{peeters2023leopardseal}         & & & & & & & & & \pcirc & \\ \hline
Marlin~\cite{tucker2025detecting}                 & & & \pcirc & & & & & & & \\ \hline
Murat~\cite{nakarmi2021murat}                     & \pcirc & & & & \pcirc & & & & & \\ \hline
Nakarmi \etal~\cite{nakarmi2022applying}          & \pcirc & & & & & & & & & \\ \hline
Park~\cite{park2023we}                            & \pcirc & \pcirc & & & \pcirc & & & & & \\ \hline
PHOENIX~\cite{echeverria2021phoenix}              & & \pcirc & \pcirc & \pcirc & & & \pcirc & & & \\ \hline
Rayhunter~\cite{rayhunter}                        & & & \pcirc & & \pcirc & \pcirc & \pcirc & & & \\ \hline
SeaGlass~\cite{ney2017seaglass}                   & \pcirc & & \pcirc & & \pcirc & \pcirc & \pcirc & & & \\ \hline
Steig \etal~\cite{steig2016network}               & & & & & \pcirc & & & & & \\
\midrule
\rowcolor{gray!15}
\textbf{Operational Variants}                     & \textbf{V3} & \textbf{V3} & \textbf{V4} & \textbf{V6} & \textbf{V1} & \textbf{V5} & $\ast$ & \textbf{V2} & \textbf{V7} & \textbf{V7} \\
\bottomrule
\end{tabular}%
}
{\raggedright \scriptsize
$\ast$~Null ciphering can be covered by V5. Rather than triggering easily observable protocol anomalies (null cipher use in LTE/5G), an adversary without security keys can alternatively exploit legacy networks via standard-compliant redirection, where encryption is inherently weak or absent (2G/3G).\par}
\vspace{-5mm}
\end{table}

We first structure the detection taxonomy and analyze the underlying primitives~(\autoref{tab:fbs_detection_primitives}; refer to \autoref{appendix:detection} for detail) as an indexing lens to bound the search space, since exhaustively searching the vast specifications is intractable. This lens identifies standardized artifacts monitored by existing detectors to pinpoint relevant specification clauses and map them onto feature dimensions, rather than introducing detector-specific heuristics. We then derive \devil's operational variants across seven dimensions~(\autoref{tab:adversarial_mapping_vertical}) as specification-permitted alternatives within these clauses. Accordingly, these variants characterize the design space that realistic adversaries may occupy within specification constraints. Coverage gaps in~\autoref{s:evaluation} reflect assumption-bounded design and evaluation, rather than targeted evasions.

\autoref{tab:fbs_detection_primitives} summarizes 10 detection primitives identified across 22 existing detection systems
-- standardized artifacts to distinguish FBSs from legitimate infrastructure -- and shows operational coverage of variants across primitives.

While V2, V5, and V7 are first introduced to serve as FBS operations, others (though previously discussed in theory) are first designed and integrated as functional variants into an end-to-end system by \devil.

\nibf{V1) Environment adaptation.}
Rather than fixed or randomized configurations, C-FBS supports adapting broadcast identifiers~\cite{3gpp_36331} based on reconnaissance. \devil extends this into fine-grained dynamic adaptation through three strategies: (1) full parameter alignment (\textit{full-adaption}) -- replicating collected identifiers, radio configurations, and reselection properties to match observed network environment, (2) \textit{Paging reproduction} -- matching timing, structure, and content of legitimate paging broadcasts, and (3) controlled parameter variation (\textit{UE-recovery}) -- selective TAC modification to enable post-attack UE recovery to legitimate networks while preserving other parameters~\cite{3gpp_24301}. C-FBS implements limited identifier configuration; \devil extends this through granular parameter control, enabling environment-specific adaptation.

\nibf{V2) TA diversification.}
3GPP TS 36.321~\cite{3gpp_36321} specifies TA commands in random access, with values 0-1,282 representing propagation delay compensation (each unit: 78m). 

In our empirical observation, C-FBS typically exhibits limited geographic coverage with a distribution of lower TA values ($\leq$ 1) due to transmission power constraints.
Meanwhile, legitimate networks serving larger geographic areas produce distributed TA measurements. Realistic adversaries with constrained coverage may diversify TA distributions to reflect broader operational context. \devil models TA diversification by randomizing values within 0-30 range -- we empirically validated that it does not impair UE connectivity while producing geographic distribution diversity.

\nibf{V3) Power-efficient hijacking.}
Although C-FBS allows for reduced transmission power to enhance stealth and power efficiency, achieving hijacking with constrained power requires extensions grounded in specification-informed logic; 3GPP standards define multiple mobility management mechanisms -- handover for connected UEs~\cite{3gpp_36331} and cell reselection for idle UEs~\cite{3gpp_36304}. 
By exploiting them, \devil can succeed with modest relative power differences to effectively take over UE connections in hijacking phase (\autoref{appendix:eval_efficacy} and~\autoref{fig:connection_hij_flow}).

\devil manipulates the \textit{handover} process~\cite{3gpp_36331} by emulating legitimate neighbors through PCI reuse and transmitting stronger signals, influencing the UE’s \tts{Measurement Report}~\cite{bitsikas2021don,shaik2018impact,samin2024gotta}. The \tts{Measurement Report}, including the signal strength and quality of neighboring cells, is crucial for triggering handover decisions in the network. For instance, when a neighboring cell’s signal meets the measurement criteria and exceeds the threshold configured by the serving network, the UE triggers a \tts{Measurement Report} (\tts{Event A4}) to the network, which then decides to initiate a handover~\cite{3gpp_36331,bitsikas2021don}. By manipulating this report, \devil can induce the network to initiate a handover through an \tts{RRC Connection Reconfiguration}.

Although this procedure fails on \devil's side due to the lack of legitimate UE context (\ie, handover failure), the UE eventually connects to \devil by redirecting to the strongest available cell, completing the hijacking.  
This mechanism is only effective for connected mode UEs, yet requires a lower power advantage ($>$ 3\,dB) than jamming.

To target idle mode UEs with lower transmission power, \devil supports \textit{cell reselection} based hijacking~\cite{heijligenberg2024attacks}.
When in idle mode, UEs periodically evaluate surrounding cells and reselect the serving cell based on signal quality and reselection criteria broadcast in SIB~\cite{3gpp_36304}. These criteria, which vary depending on intra-frequency or inter-frequency, include signal thresholds and frequency priorities.

\devil can exploit this mechanism by configuring its operating frequency band to have the highest reselection priority and broadcast stronger signals ($>$ 5\,dB). Since SIB messages are transmitted in plaintext, reselection priorities can be identified, allowing \devil to adjust its settings accordingly. Through this mechanism, \devil increases the likelihood that idle UEs will reselect to it.

Furthermore, connection hijacking often induces abnormal failures, such as RLFs or handover failures. These disruptions leave observable anomalies in the form of RLF reports or \tts{RRC Reestablishment Requests}. 
With cell reselection, \devil can redirect UEs without generating such RRC failures.
Although UE states cannot be directly controlled, the cellular nature of frequent transitions between idle and connected modes allows \devil to capture targets while reducing both signal and failure anomalies.

\nibf{V4) Alternative/Adaptive identity collection.}
While C-FBS adopts an aggressive volume of \tts{Identity Request} for IMSI-catching, realistic adversaries may employ specification-compliant alternatives and adaptive application strategies. 3GPP TS 24.301~\cite{3gpp_24301} and a prior study~\cite{tucker2025detecting} specify multiple ways of triggering identity exposure. 

\devil models identity collection diversity: (1) \textit{alternative signaling (Reject-based IMSI catching)} -- sending crafted \textit{NAS Reject} with specific causes (\eg, \#9 \textit{UE identity cannot be derived by the network}) instead of issuing explicit \tts{Identity Request}, (2) \textit{adaptive sampling} -- arbitrarily or selectively collecting identities from a fraction of UEs (\eg, 10–20\%) rather than universal, and (3) \textit{targeted operation} -- IMSI/GUTI-based victim filtering for focused collection. Here, (2) trades catch rate for stealth against volume-anomaly counters, while (3) models targeted tracking under prior knowledge.
\devil's reconfigurability also extends adaptive and targeted operations to other applications.

\nibf{V5) Alternative redirection.}
3GPP TS 36.331~\cite{3gpp_36331} specifies \tts{idleModeMobilityControlInfo} as an optional field in \tts{RRC Connection Release}, designed to provide dedicated cell reselection priorities upon radio connection termination.
This mechanism serves as a further redirection vector beyond conventional methods utilized by C-FBS and actively monitored by detectors -- SIB7 for cell reselection guidance~\cite{rayhunter,apple_patent} or \tts{redirectedCarrierInfo}~\cite{rayhunter}. 

\devil models \tts{idleModeMobilityControlInfo}-based redirection by configuring the highest reselection priority for target legacy frequencies in \tts{RRC Connection Release}, enabling UE transition to \devil's concurrent 2G/3G cells through standard reselection procedures. This specification-compliant alternative is featured alongside conventional mechanisms (SIB7, \tts{redirectedCarrierInfo}), demonstrating protocol flexibility for legacy network transitions. C-FBS implements conventional redirection with \tts{redirectedCarrierInfo}; \devil extends coverage to alternatives beyond fixed commercial implementations.

\nibf{V6) Reshaping message sequence.}
Realistic adversaries may diversify protocol message sequences based on operational context. Representatively, although C-FBS employs \textit{NAS Reject}~\cite{3gpp_24301} for connection termination as a post-attack operation, it can be replaced with \tts{RRC Connection Release}~\cite{3gpp_36331}. They are connection termination mechanisms defined in the 3GPP specifications across different layers.

\devil models this type of sequence reshaping through connection management flexibility (\textit{Exclude REJECT}): (1) \tts{RRC Connection Release} as a specification-driven alternative to \textit{NAS Reject}, terminating connections at radio layer~\cite{park2023we} or (2) silent -- connection timeout through non-response to UE signaling~\cite{paci2025flashcatch}. These alternatives reflect protocol-layer flexibility enabling sequence variation.

\nibf{V7) Hardware compensation.}
3GPP specifications~\cite{3gpp_36401,3gpp_38401} define synchronization features for physical-layer signal integrity -- encompassing both frequency stability and timing accuracy -- to prevent inter-cell interference and maintain network stability. Legitimate infrastructure typically utilizes high-precision synchronization (from external clocks or operator-provided sources) and calibrated radio front-ends.
In contrast, adversaries often rely on low-cost or resource-constrained hardware, \eg, the bandwidth-restricted SDR utilized by C-FBS. Such hardware imperfections manifest as observable physical-layer artifacts, including timing misalignments in transmitted radio frames~\cite{apple_patent}, center frequency offsets (CFO), and modulation distortions~\cite{zhuang2018fbsleuth}.

\devil models physical-layer diversity through hardware compensation strategies: (1) \textit{Time synchronization}: employing external timing sources (\eg, GPSDO) to align frame timing with sub-microsecond accuracy and stabilize the internal clock, effectively mitigating timing errors and addressing hardware-originated artifacts, and (2) \textit{RF tuning}: adjusting RF outputs and reducing modulation imperfections by injecting controlled CFO or shaping pilot signal magnitudes to reflect the RF characteristics of legitimate cells or represent the breadth of diverse hardware, ranging from entry-level SDRs to high-end systems (\$1M~\cite{css_jacob_brochure}).

\subsection{Dependency Checker}
\label{ss:dep_checker}
\begin{figure}[h]
    \centering
    \includegraphics[width=\linewidth]{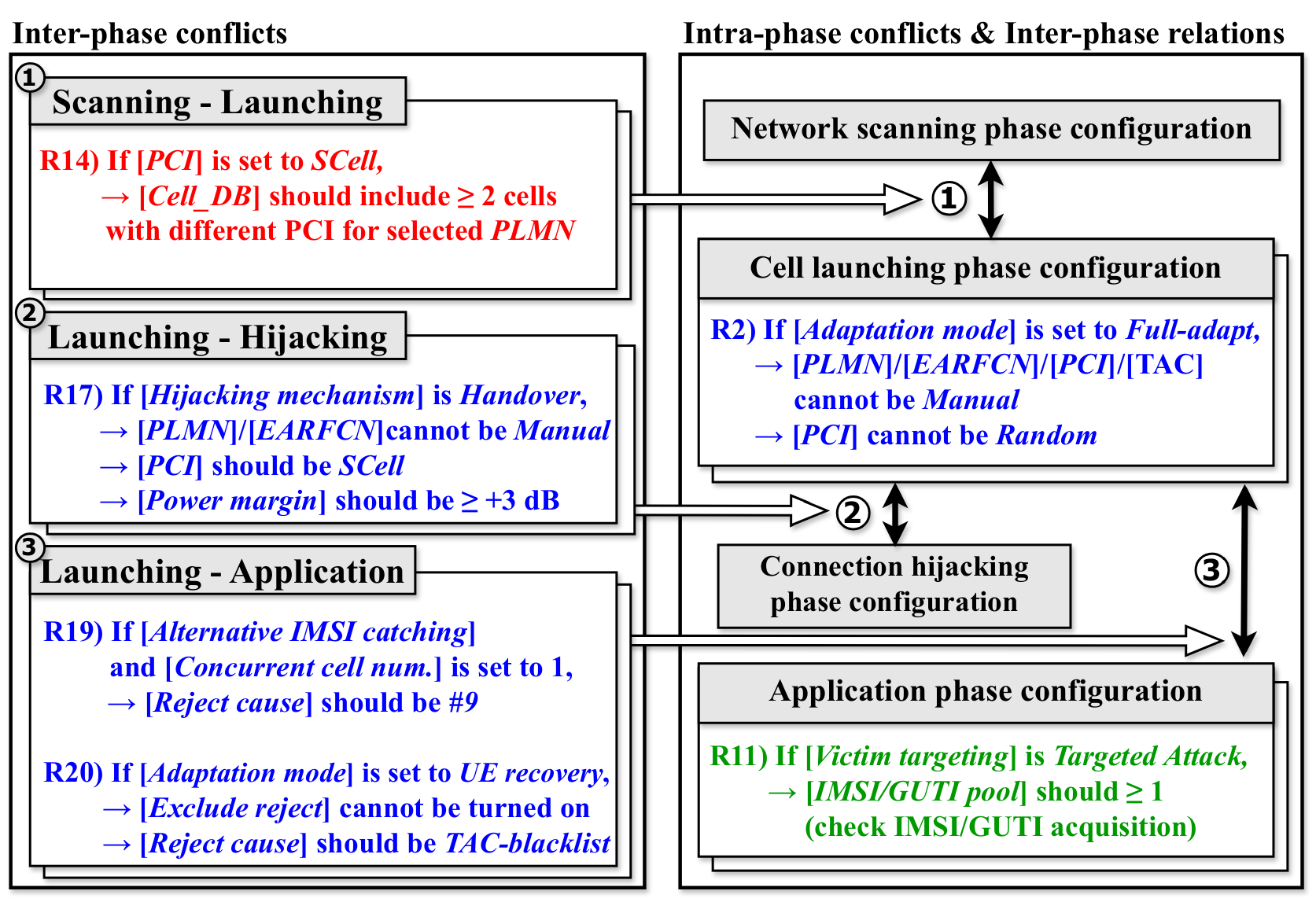}
        \caption{
        Dependency checker with example rules. Arrows denote inter-phase relations. Colors indicate resolution types: automatic (blue), strict validation (red), prerequisite enforcement (green).}
    \label{fig:dependency_checker}
\end{figure}

As indiscriminate exploration of the configuration space across the four-phase operational pipeline and adversarial variants risks yielding logical conflicts, \devil integrates an editable, YAML-formatted rule-based dependency checker to prune invalid configuration sets. Specifically, it filters out both explicit 3GPP standard violations and empirically ineffective operations, with the latter validated using commercial UEs and networks. As exemplified in~\autoref{fig:dependency_checker} (Appendix), these dependencies are classified into two types; intra-phase dependencies involve parameter constraints within a single phase (\textit{Rule 2 and 11}), while inter-phase dependencies span multiple phases where former phase configurations impose constraints on subsequent phases (\textit{Rule 14, 17, 19, and 20}). The full set of 21 dependency rules is available at~\cite{devilray_demo}.

These 21 rules address configuration conflicts by encoding specification and empirical constraints bounded by our operational pipeline and feature dimensions~(\autoref{ss:baseline}), with each rule justified by a cited 3GPP clause or a documented empirical observation~\cite{devilray_demo}, rather than an exhaustive search. 

\nibf{Specification-based rules.}
We define nine specification-based rules to capture configuration constraints implied by 3GPP specifications. For example, if the hijacking mechanism is set to handover (V3), the launched cell cannot be configured with arbitrary identifiers: its PLMN and EARFCN should be derived from scanned network information, its PCI should be selected from a scanned neighboring cell, and its power margin should be sufficient to trigger UE measurement reporting (\textit{Rule 17}). 

\nibf{Empirically grounded rules.}
To handle practical constraints beyond standards-compliance yet essential for operational effectiveness, we introduce 12 empirically grounded rules as a complement. A representative case is \textit{Rule~19}, which governs \textit{NAS Reject} cause selection for IMSI-catching.
Although multiple reject causes are available for alternative IMSI-catching (V4)~\cite{tucker2025detecting}, certain causes force the UE to blacklist the current PLMN/TAC and attempt an IMSI Attach to alternative networks, which is incompatible if \devil operates with a single-cell. As this behavior is inherently UE-dependent, we tested it on six COTS UEs (three baseband vendors) against all causes~\cite{tucker2025detecting} and found that cause \#9 is the only option that consistently achieves the intended identity exposure in a single-cell setting.

\nibf{Conflict resolution.}
\devil's dependency checker semantically identifies rule violations at runtime and automatically addresses them through one of the following ways:
\begin{itemize}
    \item \textbf{Automatic Resolution}: It adjusts former phase configurations to satisfy latter phase one to ensure operational feasibility, \eg, filtering PCI candidates from the scanning database to match selected hijacking strategies (\textit{Rule 17}).
    \item \textbf{Strict Validation}: Unresolvable conflicts, such as configuring handover but neighbor cells not found, trigger configuration rejection with diagnostic output (\textit{Rule 14}).
    \item \textbf{Prerequisite Enforcement}:
    It rejects configurations lacking operational prerequisites and specifies corresponding requirements -- ensuring that IMSI/GUTI acquisition precedes targeted attacks (\textit{Rule 11}) and validating redirection parameters match configured 2G carrier. 
\end{itemize}

\subsection{Reconfigurable Implementation}
\label{ss:implementation}
We build \devil as a reconfigurable system by extending srsRAN~\cite{srsran}, OpenBTS~\cite{openbts}, and LTESniffer~\cite{hoang2023ltesniffer} with 8K lines of code. It supports diverse single- to multi-SDR hardware setups, including USRP B210s~\cite{usrp_b210} and high-end X310s~\cite{usrp_x310}, for scalable multi-cell combinations. 

\begin{figure*}[t]
    \centering
    \includegraphics[width=0.8\linewidth]{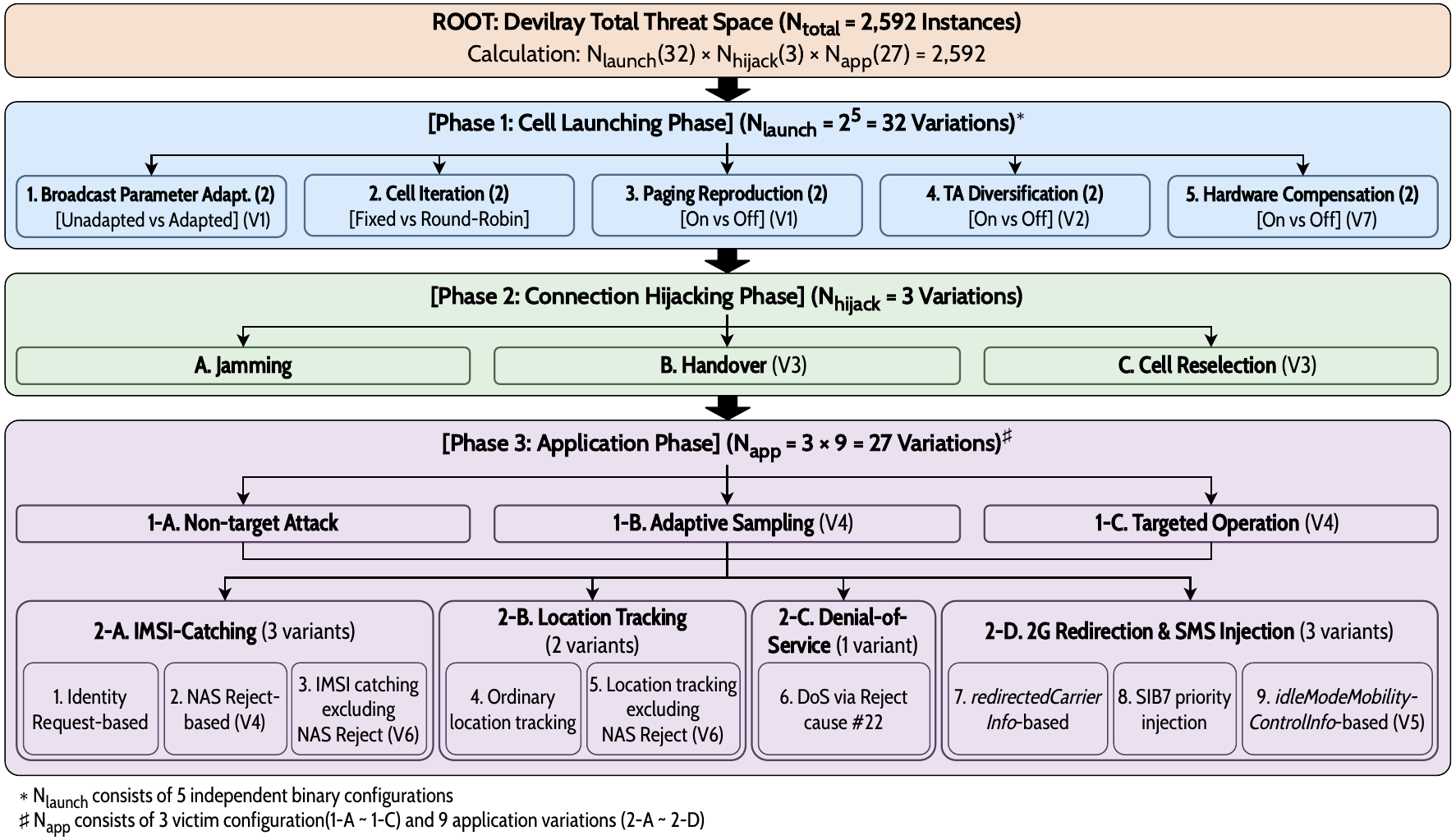}
        \caption{\devil's configurations space to explore FBS instances}
    \label{fig:devilray_instance_tree}
\end{figure*}

\devil enables the logical exploration of 2,592 adversarial instance sets. Out of the 55K product space from~\autoref{fig:devilray_config}, the dependency checker filters 46K invalid configurations, yielding 9K valid ones. To prevent artificial inflation, we conservatively exclude per-field permutations within each logical combination, resulting in 2,592 instances.
This 2,592 count represents the logically explorable state space, rather than an arbitrary experimental tally, bounded by our encoded rules and operational dimensions, serving as a structured domain for detection diagnostics.
Unique instance derivation is detailed in~\autoref{appendix:numbers}, while the configuration space is illustrated in~\autoref{fig:devilray_instance_tree} (Appendix).

Rather than claiming attack efficacy (assessed in~\autoref{s:evaluation}) for every configuration, we provide empirical evidence of feasibility across three scopes: (1)~\textit{per-phase~operability}, where we verified all constituent behaviors (32 launching, 3 hijacking, 27 application); (2)~\textit{end-to-end testing}, where we instantiated and validated all 2,592 checker-accepted configurations through interaction with UEs to verify that each completes; and (3)~\textit{negative validation}, where we deliberately executed rule-violating configurations (\eg, absent neighbor cells for handover, \textit{Rule 14}) and observed failures as predicted, confirming correct pruning.

Complete end-to-end configurations can be exported as reusable YAML-based profiles, released alongside our implementation and packet captures as a reproducible reference platform. To help researchers navigate this space, \devil structures these profiles into operational families (\eg, silent hijacking and high-false-positive induction), providing clear entry points for detector stress-testing.
  \section{Evaluation}
\label{s:evaluation}

We examine security risks (\ie, blind spots) in existing FBS detection mechanisms (\textbf{RQ4}) while demonstrating \devil's utility by evaluating seven detectors -- currently accessible in public and maintained, yet covering diverse primitives~(\autoref{tab:fbs_detection_primitives}) -- under realistic, specification-driven adversarial instances configured by \devil\footnote{Prior to this evaluation, we validate \devil's effectiveness on connection hijacking and application across seven COTS UEs in~\autoref{appendix:eval_efficacy}.}.
We first evaluate currently accessible detectors against IMSI-catching, a representative protocol-layer workflow in~\autoref{ss:eval_showcase1}, and then extend the evaluation to physical-layer detection primitives in~\autoref{ss:eval_showcase2}. Our setup consists of a laptop running \devil software connected to two B210s.

\subsection{IMSI-catcher Instances}
\label{ss:eval_showcase1}

We first evaluate seven publicly obtainable detection systems~\cite{echeverria2021phoenix,arnold2024catch,crocodilehunter,rayhunter,eaglesecurity,samin2024gotta,tucker2025detecting}\footnote{IMSI-catcher-catcher~\cite{dabrowski2014imsi} and Seaglass~\cite{ney2017seaglass} are excluded due to a decade-long lack of updates, yet are largely integrated into recent detectors as pioneering works, according to our detection primitive analysis~(\autoref{tab:fbs_detection_primitives}.)} against \devil's IMSI-catcher instances.
We prioritize IMSI-catching over other protocol-layer attacks as it remains a primary FBS operation for surveillance and is commonly addressed by the target detectors.

Rather than attempting an exhaustive physical census of all 2,592 states, we construct a targeted diagnostic sample of 10 instances by mapping each target detector's detection domain to corresponding variants -- the primary utility of \devil. This selection demonstrates how researchers can effectively narrow the configuration space for their detector diagnostics.
Specifically, we aggregate target systems' detection primitives~(\autoref{tab:fbs_detection_primitives}) into three domains -- cell parameters, hijacking, and IMSI-catching methods -- and sample instances using domain-relevant variants (V1, V3, V4, V6).
The environment adaptation variant (V1) serves as a baseline parameter with two binary settings: \textit{NA} (adaptation disabled) and \textit{A} (enabled).
While using \textit{NA} to establish a naive baseline for initial comparison, we explore the impact of hijacking (V3) and application-phase variations (V4, V6) by fixing V1 to \textit{A} and performing a $3 \times 3$ cross-product sweep of three hijacking mechanisms (\textit{J, H, C}) and three application-phase variations (\textit{I}: \tts{Identity Request} then \textit{NAS Reject}, \textit{IR}: Reject-based catching, \textit{Ir}: \tts{Identity Request} then RRC release). This yields 10 instances, as shown in~\autoref{tab:devil_evasion_eval}. Adaptive IMSI-catching (V4) is examined separately against the statistical approach~\cite{tucker2025detecting} later.

\begin{table}[t]
\centering
\caption{Evaluation results of IMSI-catcher instances. \devil identifies adversarial blind spots in 6 detectors.}
\label{tab:devil_evasion_eval}
\setlength{\tabcolsep}{2.5pt}
\resizebox{\linewidth}{!}{%
\begin{tabular}{lcccccc}
\hline
\textbf{\makecell[b]{Detection\\System}} & 
\textbf{\makecell[b]{PHOENIX$^{\ast}$\\\cite{echeverria2021phoenix}}} & 
\textbf{\makecell[b]{Rayhunter$^{\dagger}$\\\cite{rayhunter}}} & 
\textbf{\makecell[b]{CellGuard$^{\ddagger}$\\\cite{arnold2024catch}}} & 
\textbf{\makecell[b]{Crocodile\\Hunter$^{\ddagger}$~\cite{crocodilehunter}}} & 
\textbf{\makecell[b]{EAGLE\\Security$^{\ddagger}$~\cite{eaglesecurity}}} & 
\textbf{\makecell[b]{FBSDetector$^{\sharp}$\\\cite{samin2024gotta}}} \\
\hline
C-FBS                    & \fullcirc  & \fullcirc  & \fullcirc~(51)  & \emptycirc~(13) & \emptycirc~(10) & 95.6\% / -- \\
\hline
\textit{J + I + NA}     & \fullcirc  & \fullcirc  & \halfcirc~(50)  & \fullcirc~(21)  & \fullcirc~(35)  & 97.0\% / 47.8\% \\
\textit{J + I + A}      & \fullcirc  & \fullcirc  & \halfcirc~(30)  & \fullcirc~(20)  & \halfcirc~(25)  & 16.4\% / 44.6\% \\
\textit{J + IR + A}     & \fullcirc  & \emptycirc & \halfcirc~(30)  & \fullcirc~(21)  & \halfcirc~(25)  & 9.3\% / 45.7\% \\
\textit{J + Ir + A}     & \emptycirc & \fullcirc  & \emptycirc~(0)   & \fullcirc~(21)  & \halfcirc~(25)  & 73.9\% / 82.3\% \\
\hdashline
\textit{H + I + A}      & \fullcirc  & --         & \halfcirc~(30)  & \emptycirc~(5)  & \emptycirc~(10) & 95.9\% / 43.7\% \\
\textit{H + IR + A}     & \fullcirc  & --         & \halfcirc~(30)  & \emptycirc~(5)  & \emptycirc~(10) & 20.2\% / 46.8\% \\
\textit{H + Ir + A}     & \emptycirc & --         & \emptycirc~(0)   & \emptycirc~(4)  & \emptycirc~(10) & 31.7\% / 72.5\% \\
\hdashline
\textit{C + I + A}      & \fullcirc  & --         & \halfcirc~(30)  & \emptycirc~(6)  & \emptycirc~(10) & 95.9\% / 61.5\% \\
\textit{C + IR + A}     & \fullcirc  & --         & \halfcirc~(30)  & \emptycirc~(7)  & \emptycirc~(10) & 9.9\% / 47.0\% \\
\textit{C + Ir + A}     & \emptycirc & --         & \emptycirc~(0)   & \emptycirc~(4)  & \emptycirc~(10) & 30.6\% / 74.7\% \\
\hline
\end{tabular}%
}
\begin{tablenotes}
    \scriptsize
    \begin{minipage}{\columnwidth}
        \item[] \scalebox{0.7}{\fullcirc}: Detected, \scalebox{0.7}{\halfcirc}: Suspect, \scalebox{0.7}{\emptycirc}: Missed. Detection score given in parentheses~($\ddagger$ below).
        \item[] \textit{J}: Jamming, \textit{H}: Handover, \textit{C}: Cell reselection, \textit{I}: IMSI-catching with Identity Request, \textit{IR}: IMSI-catching with \textit{NAS Reject}, \textit{Ir}: IMSI-catching with Identity Request then RRC release, \textit{NA}: Non-Adaptation, \textit{A}: Adaptation. 
        \item[] $\ast$ Evaluated via the PHOENIX implementation in FBSDetector~\cite{samin2024gotta}, as the original system is not publicly available.
        \item[] $\dagger$ Rayhunter requires an Orbic mobile hotspot~\cite{orbic}, which failed mobile data attachment locally, precluding handover and cell reselection. Nevertheless, jamming fully reflects Rayhunter's message-based detection, and handover and reselection are expected to yield the same results.
        \item[] $\ddagger$ Suspiciousness scores (0-100 scale): CellGuard (0-5 trusted, 6-50 anomalous, $\ge$51 suspicious), Crocodile Hunter ($\ge$20 suspicious), and EAGLE Security (0-19 trusted, 20-25 anomalous, $\ge$26 suspicious).
        \item[] $\sharp$ Detection rate on NAS-layer (left) and RRC-layer (right) models, respectively. For C-FBS, only NAS detection is available, as it lacks compatible RRC packet captures.
    \end{minipage}
\end{tablenotes}
\vspace{-5mm}
\end{table}

\nibf{Detection results on 10 instances.}
\autoref{tab:devil_evasion_eval} presents evaluation outcomes for six detectors 
using the 10-instance set, comparing with C-FBS. Across all six detectors, \autoref{tab:devil_evasion_eval} reveals at least one instance where detection fails, though no single instance is universally missed by all systems at once. We conduct a root cause analysis to identify the blind spots rooted in assumption-bound designs.

PHOENIX~\cite{echeverria2021phoenix} operates under the rigid assumption that adversarial connection termination invariably relies on \textit{NAS Reject} messages. This dependency creates a structural blind spot; because it monitors only the NAS layer for termination, it misses the \textit{Ir} instances, which shift termination to the RRC layer (\tts{RRC Connection Release}).

Rayhunter~\cite{rayhunter} exhibits a gap rooted in an assumption-bound design that posits identity disclosure is driven exclusively by explicit \tts{Identity Request} or \textit{IMSI paging}. This heuristic creates a blind spot against the \textit{IR} instance utilizing alternative signaling (\textit{NAS Reject}) to capture IMSIs.

CellGuard~\cite{arnold2024catch} demonstrates how heuristic-based scoring introduces blind spots when encountering flexible operational dimensions. Its design assumes static signatures, assigning fixed weights to \textit{NAS Reject} presence (30 points) and location database anomalies (20 points). While the naive \textit{J+I+NA} instance triggers these rules, \devil's environment adaptation variants (\textit{A}) mimic legitimate cell attributes, leaving the check ineffective. Furthermore, its high-power threshold is not triggered by all instances due to an excessively high threshold. Consequently, all \textit{Ir+A} instances are unflagged with zero points.

Crocodile Hunter~\cite{crocodilehunter} reflects detection failures tied to static environmental assumptions, relying heavily on an external database (OpenCelliD~\cite{opencellid}) and high-power signatures. In our evaluation, when OpenCelliD lookups failed due to cell tracking or update issues, its benign-by-default handling left both unadapted and adapted instances unflagged. Moreover, as its heuristics focus on high-power profiles, \devil's power-efficient variants (\textit{H} and \textit{C}) stay below its suspiciousness threshold (20).

EAGLE Security~\cite{eaglesecurity} also exhibits database-dependent behavior. As the PLMN used in our experiments is absent from its internal database, it assigns a baseline score of 10 to all variants. For the \textit{NA} instance, it adds 10 points, marking cell information ``UNKNOWN''. Jamming cases cause the observation of a single searchable cell, adding 15 points.

Finally, the ML-based FBSDetector~\cite{samin2024gotta} shows detection rates below its reported performance (over 90\%) for specific variants across 1,000 test trials each. We observe a layer-dependent trend rooted in their assumption-bound designs: the NAS-layer model often misses \textit{IR} instances by overlooking alternative signaling, while the RRC-layer model detects \textit{Ir} instances better due to a design involving \tts{RRC Connection Release} based exploitations. Since the root cause of an ML model is not readily interpretable, we instead retrain it on a dataset including \devil's variants to examine detector improvement and trade-offs.

Specifically, retraining the model using \devil-identified failure cases featuring jamming and handover~(\autoref{tab:devil_evasion_eval}) demonstrates performance gains on unseen \devil instances featuring cell reselection, with average true positive rates (TPR) improving from 27\% to 63.3\% (NAS layer) and from 50.8\% to 95.8\% (RRC layer).
Concurrently, average false positive rates (FPR) increased from 4.7\% to 8.8\% and from 1.4\% to 5.3\%, respectively. Notably, retraining specifically on the lowest-performing instance (\textit{C+IR+A}) causes the FPR to reach 15\% (NAS). These results suggest that simply expanding a detector's decision boundary incurs practical trade-offs in deployment.

\nibf{Adaptive IMSI-catching.}
\begin{figure}[t]
    \centering
    \includegraphics[width=0.9\columnwidth]{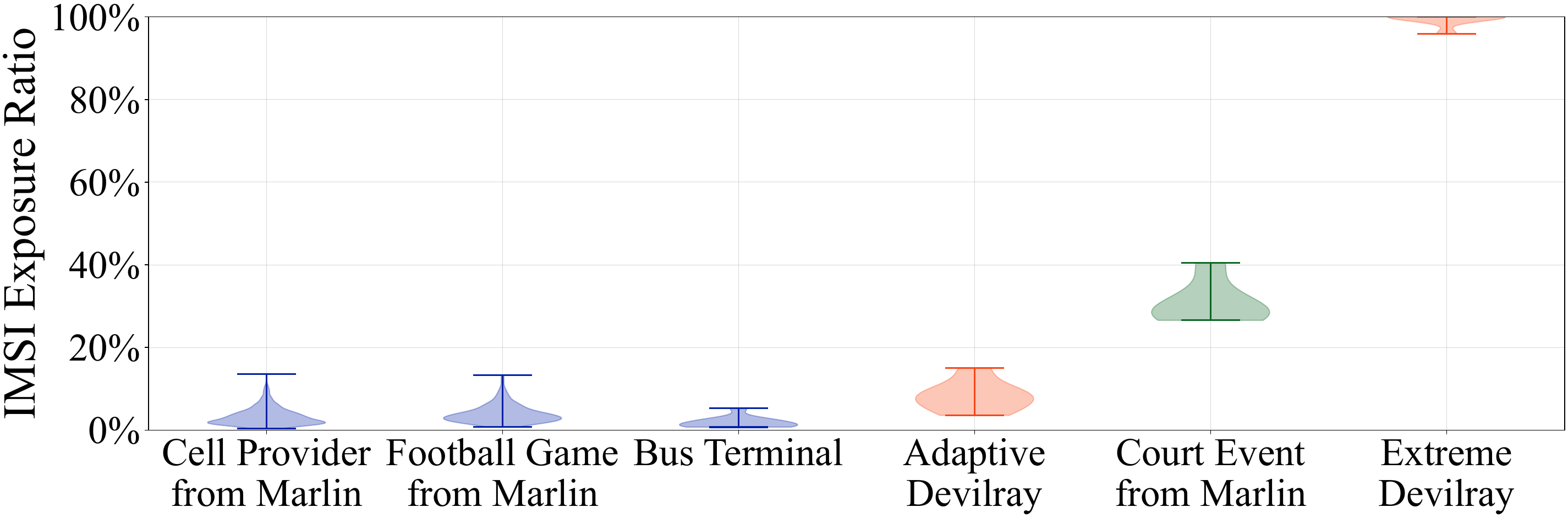}
    \caption{Distribution of IMSI-exposing messages. It shows the impact of \devil's variation on statistical significance. \vspace{-5mm}
    }
    \label{fig:adaptive_attack_figure}
\end{figure}
We evaluate Marlin~\cite{tucker2025detecting}, which performs statistical monitoring of IMSI-exposing messages, by comparing extreme (100\% exposure) and adaptive (10\%, V4) \devil instances. Using five COTS UEs and 50 trials per instance with LTESniffer~\cite{hoang2023ltesniffer}, we find that adaptive \devil produces a distribution centered around 10\% -- overlapping with benign environments in Marlin’s dataset and our bus-terminal captures -- whereas the extreme instance approaches 100\% (\autoref{fig:adaptive_attack_figure}). This suggests that Marlin’s statistical logic is sensitive to the aggregate rate of IMSI-exposing messages; by moderating this rate through adaptive collection, \devil shifts its profile into the benign range, affecting detection logic tuned for aggressive behaviors.

These results demonstrate that narrow and fragmented threat models in current detection mechanisms lead to adversarial blind spots on different realistic threats. 
By leveraging \devil to relax these rigid threat assumptions, we show that even established detection logic struggles to generalize to the realistic diversity of specification-driven operations.

\subsection{Physical-layer Instances}
\label{ss:eval_showcase2}

We now evaluate physical-layer detection with \devil under hardware compensation (V7).

\begin{figure}[t]
    \centering
    \includegraphics[width=.9\columnwidth]{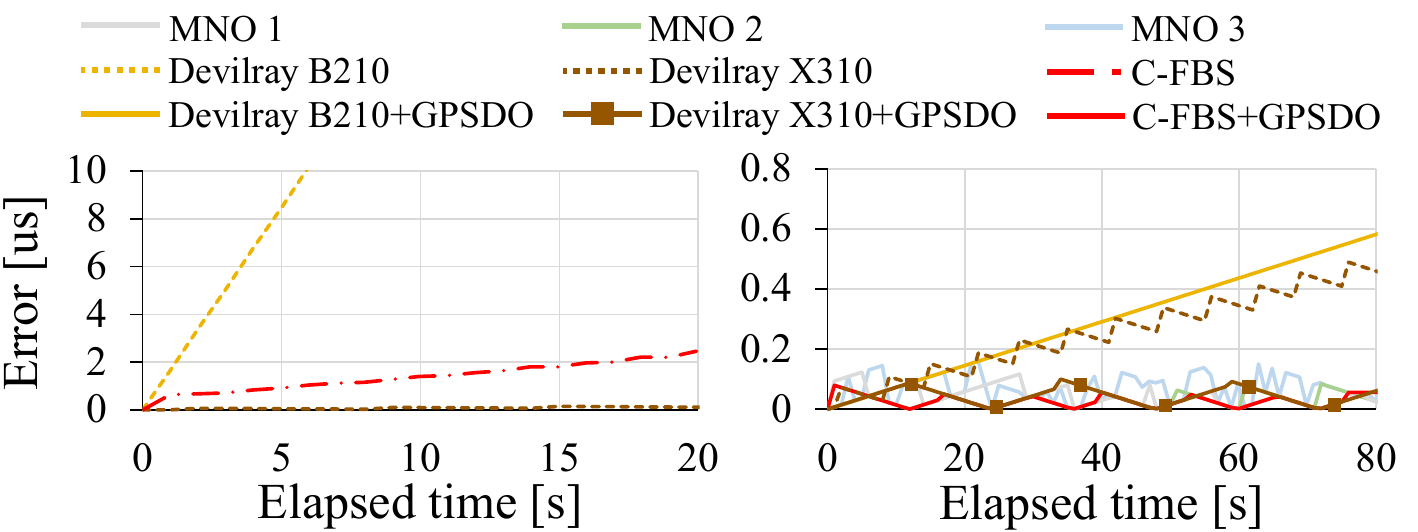}
    \caption{
    Cumulative timing misalignment of radio frames. Note the different axis scales to highlight the magnitude of misalignment and the precision of compensation. \vspace{-3mm}
    }
    \label{fig:time_sync_error}
\end{figure}

\nibf{Transmission timing error.}
To evaluate the robustness of transmission accuracy primitives against V7, we measure timing misalignment between received radio frames using LTESniffer~\cite{hoang2023ltesniffer} and track cumulative drift over time. Specifically, we compute per-frame timing error and compare cumulative error across benign cells from three MNOs, \devil (B210 and X310), and C-FBS. To assess the effect of external clock, we include measurements for GPSDO-equipped setups. 

\autoref{fig:time_sync_error} presents the cumulative timing error across different cells. While the B210 and C-FBS without GPSDO exhibit substantial cumulative drift (left plot), the right plot reveals that equipping the C-FBS and X310 with a GPSDO reduces cumulative drift to sub-microsecond ranges. This hardware compensation (V7) renders \devil nearly indistinguishable from legitimate infrastructure.

\nibf{RF characteristics.}
\begin{figure}[t]
    \centering
    \includegraphics[width=.9\columnwidth]{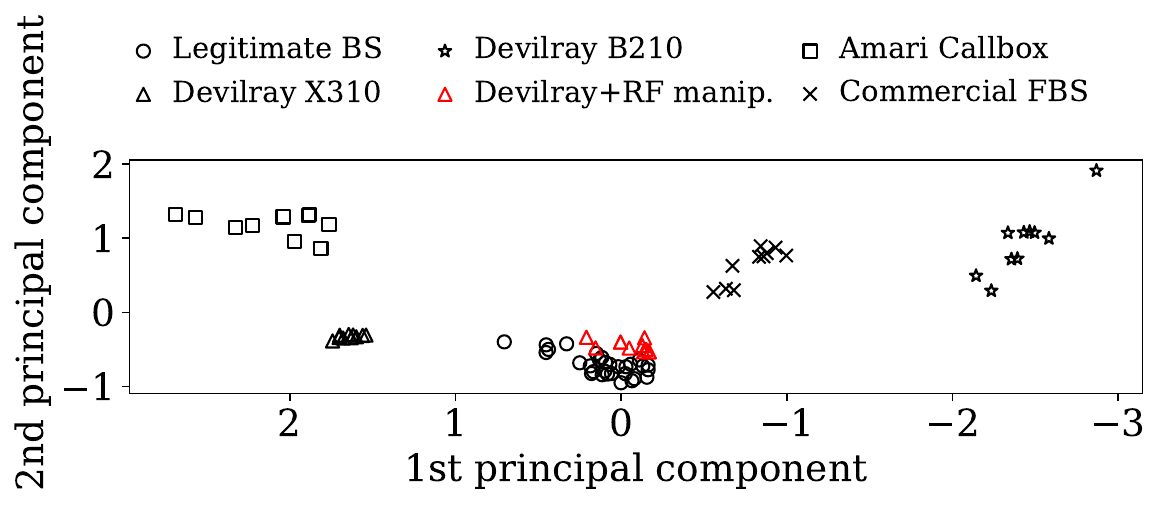}
    \caption{PCA results of RF features. It highlights \devil successfully modulates its hardware profile to align with legitimate cells via RF tuning.\vspace{-5mm}}
    \label{fig:pca_figure}
\end{figure}
We extend RF fingerprinting -- originally proposed by FBSleuth~\cite{zhuang2018fbsleuth} for 2G -- to LTE and evaluate it under hardware compensation (V7). We utilize three features: CFO, time synchronization error, and magnitude error~(\autoref{appendix:rf} for detail). To examine their effectiveness, we measured these features across three MNOs, \devil, Amari Callbox~\cite{amari}, and C-FBS. While they initially exhibit distinct patterns that differentiate them from legitimate cells, we further explore the impact of hardware compensation.

Using Principal Component Analysis (PCA), we visualize the projection space defined by legitimate cells and uncompensated FBS instances. As shown in~\autoref{fig:pca_figure}, legitimate cells form tight clusters, while uncompensated FBS devices form separate, distinguishable clusters. However, by applying RF manipulation on \devil (V7), which introduces controlled software-based CFO injection, the resulting features become indistinguishable from legitimate cells. 

These results reveal blind spots in detection logic that assumes FBS operations are inherently limited by hardware imperfections; we demonstrate that operational variants involving hardware compensation (V7) in \devil leads to gaps for detection relying on physical-layer artifacts.

\nibf{Takeaway.} 
Our evaluation shows that current detection mechanisms have blind spots under realistic variations of FBS behavior. While the union of the evaluated detectors covers all tested configurations, this does not directly imply a practical combined detector. 
If alarms are simply combined with an OR rule, the union FPR under benign traffic $B$ satisfies $\mathrm{FPR}_{\mathrm{union}} = P(\bigcup_i F_i \mid B) \ge \max_i \mathrm{FPR}_i$ (where $F_i$ is detector $i$'s false alarm event). Thus, combining detectors cannot lower the highest individual FPR and can further increase when different detectors falsely flag different benign samples.
More sophisticated integration may behave differently, but would require a new detector design and separate evaluation. Additionally, integrating heterogeneous detectors across diverse entities and vantage points (raw IQ, SIB broadcasts, OS and connection logs) may introduce runtime synchronization overhead and race conditions.

Our FBSDetector experiment shows that improving attack coverage can incur a substantial false-positive cost in practice. After retraining with previously missed Devilray instances, the average TPR improves, while the corresponding FPR increases. When retrained specifically to cover the hardest missed instance (\textit{C+IR+A}), the FPR reaches 15\%, meaning that roughly one in seven benign samples is falsely classified as FBS activity. These results show that broader attack coverage alone does not imply a deployable detector; the resulting false-positive cost must also be evaluated.
  \section{Related Work}
\label{s:related}
\begin{table}[t]
\centering
\caption{\devil vs. prior FBS implementations.}
\label{tab:relatedworks}
\setlength{\tabcolsep}{4pt}
\resizebox{.9\linewidth}{!}{%
\begin{tabular}{l cccc}
\toprule
& \textbf{\makecell[b]{Connection\\hijacking}} 
& \textbf{\makecell[b]{Variants\\(\autoref{ss:evasion})}} 
& \textbf{\makecell[b]{Reconfig.\\instances}} 
& \textbf{\makecell[b]{Artifact\\Avail.}} \\ 
\midrule
Bitsikas~\etal~\cite{bitsikas2021don}              & \textit{H}          & V3$^\ast$         & 1     & $\times$ \\ 
CellGuard~\cite{arnold2024catch}                   & \textit{J}          & --                & 1     & $\times$ \\
FBSDetector~\cite{samin2024gotta}                  & \textit{J, H}       & V1, V3$^\ast$, V6 & 5     & $\times$ \\
FlashCatch~\cite{paci2025flashcatch}               & \textit{J}          & V6                & 1     & $\times$ \\
Heijligenberg~\etal~\cite{heijligenberg2024attacks}& \textit{C}          & V3$^\ast$         & 1     & $\times$ \\
Park~\cite{park2023we}                             & \textit{J, H}       & V1, V3$^\ast$, V6 & 11    & $\times$ \\
PHOENIX~\cite{echeverria2021phoenix}               & \textit{J}          & V6                & 1     & $\times$ \\
Shaik~\etal~\cite{shaik2018impact}                 & \textit{H}          & V3$^\ast$         & 1     & $\times$ \\
\midrule
\rowcolor{gray!12}
\textbf{\devil}                                    & \textbf{\textit{J, H, C}} & \textbf{V1--V7}  & \textbf{2,592} & \ding{52} \\ 
\bottomrule
\end{tabular}%
}\par
\parbox{.9\linewidth}{\raggedright \scriptsize
\textit{J}: Jamming, \textit{H}: Handover, \textit{C}: Cell reselection.\\
$^\ast$Focusing exclusively on either handover or cell reselection.}
\vspace{-5mm}
\end{table}
As summarized in~\autoref{tab:relatedworks}, prior FBS implementations~\cite{bitsikas2021don,arnold2024catch,samin2024gotta,paci2025flashcatch,heijligenberg2024attacks,park2023we,echeverria2021phoenix,shaik2018impact} are largely restricted to static adversaries (limited instances) or isolated hijacking mechanisms. In contrast, \devil enables combinatorial configurations across three hijacking mechanisms and seven variant dimensions, exploring 2,592 instances.
Furthermore, unlike existing study-specific prototypes that remain unavailable, \devil provides an archived, reconfigurable adversarial baseline with reusable profiles to facilitate reproducible research and cross-detector benchmarking.
  \section{Limitations}
\nibf{Single C-FBS and generalizability boundaries.}
While our empirical analysis was constrained to a single C-FBS product due to restricted access, it provides an empirical anchor grounded in real-world behaviors rather than hypothetical modeling. We extend beyond the observed C-FBS using prior work and 3GPP-defined procedures, without assuming that it represents the broader commercial ecosystem. Devilray's design space does not claim exhaustive coverage of all possible 3GPP procedures or proprietary orchestration strategies. Rather than claiming an all-encompassing platform, we position Devilray as an extensible baseline that can incorporate newly emerging threat behaviors.

\nibf{Bounded methodology and rule completeness.}
Similarly, the operational variants and dependency rules are bounded by our methodology and do not capture an unbounded adversarial space. We make no claim that the rule set is sound or complete across the entire 3GPP standard, as adversaries may exploit non-standard protocol anomalies or novel cross-RAT strategies outside our encoded rules. The dependency checker serves as a tractable, specification-grounded engine for pruning invalid setups within our defined pipeline and dimensions, with extensible and editable rules. 

\nibf{State space and evaluation trade-offs.}
Our evaluation navigates practical trade-offs between the 2,592 logically explorable states and the 10 diagnostic test cases. While all 2,592 checker-accepted configurations were verified on our testbed, evaluating every state across external detectors is constrained by over-the-air measurement overhead and baseband logging latencies. We therefore focus detector evaluation on a targeted diagnostic sample of 10 instances mapped to target detectors' primitives, demonstrating \devil's key utility in enabling targeted diagnostics without brute-force exploration. Consequently, unexamined instances within the state space may harbor residual detection gaps not characterized in this work.

\nibf{Physical-layer bounds and commercial detectors.}
Our physical-layer evaluation is bounded to baseline RF features under linear PCA, acknowledging that advanced RF fingerprinting (\eg, deep-learning-based approaches~\cite{sankhe2019oracle,riyaz2018deep}) may capture non-linear analog artifacts. Furthermore, while V7 compensates for targeted impairments, it does not eliminate all hardware-inherent characteristics (\eg, SDR sampling rate limits); broader analog compensation (\eg, software resampling) remains future work. Finally, our evaluation focuses on accessible, maintained detectors. While evaluating proprietary carrier-grade and law-enforcement systems is valuable, access remains restricted by regulatory controls, though \devil serves as a shared baseline to benchmark such systems should access be granted.

\section{Discussion}
\label{s:discussion}
\nibf{How to use \devil?}
\devil facilitates effective dataset generation across its 2,592 configuration spaces, leveraging variants mapped directly to target detectors' primitives and structural adversarial scenarios (\eg, IMSI-catchers and SMS-blasters). This enables the evaluation of existing and future detectors against realistic adversaries and the pinpointing of failures, thereby guiding detector improvements, such as threshold tuning or model refinement.

\nibf{Future mitigation.}
The insights gained from \devil can guide realistic countermeasures.
Specifically, beyond consolidating individual phase capabilities, modeling explicit cross-phase dependencies reveals detection opportunities that isolated-phase detectors miss. Existing detectors often monitor an isolated operational phase (\eg, PHOENIX~\cite{echeverria2021phoenix} on NAS sequences or FBSleuth~\cite{zhuang2018fbsleuth} on RF artifacts). By modeling the end-to-end pipeline and dependencies, \devil guides defenders on which cross-layer features to monitor and at which phase and timing.

Additionally, our primitive set and detector taxonomy (\autoref{tab:fbs_detection_primitives} and~\cite{devilray_demo}) provide a foundation for a robust defense pipeline across detection, identification, and localization. For example, monitoring comprehensive protocol- and physical-layer primitives via tools like LTESniffer~\cite{hoang2023ltesniffer} can detect FBS presence -- as attackers struggle to evade all signatures simultaneously -- while identifying adversary profiles (\eg, attack type, band, cell ID). This evidence directly facilitates subsequent physical localization (\eg, via AoA~\cite{kumar2014lte,oh2024enabling} or TDoA~\cite{kotuliak2022ltrack}) to trace and confiscate active devices.

Furthermore, we suggest two key principles for future detector design. First, claims of improved coverage require evaluating both attack detection performance and false positives on benign traffic, particularly for adversarial instances that existing detectors fail to detect. In our FBSDetector~\cite{samin2024gotta} experiment, retraining on previously missed \devil instances improved detection but also increased the FPR on benign carrier traffic, reaching 15\% for the hardest missed instance. 
Second, practical deployability requires clearly defining the detection entity (\eg, UE, sniffer, or network) and its runtime telemetry, avoiding reliance on metrics unavailable at the intended deployment point.

\nibf{Extending to 5G-SA.}
Extending \devil to 5G-SA remains future work and can proceed along two directions: (1) forcing inter-RAT fallback to LTE via intentional 5G-SA interference, SIB manipulation, or NAS Reject injection~\cite{karakoc2023never}, or (2) native 5G-SA transition, which requires New Radio (NR) signaling, SUCI-catching, and wider-bandwidth SDRs.

  \section{Concluding Remarks}
\label{s:conclusion}
FBS detection research has largely relied on self-built adversaries with fixed threat assumptions, leaving realistic adversarial behaviors underexplored. We presented \devil, a reconfigurable adversarial baseline anchored in empirical C-FBS analysis and extended via four-phase operational framework, seven feature dimensions, and specification-permitted variants. Within this methodology-bounded space, \devil can instantiate 2,592 configurations to diagnose adversarial blind spots across diverse FBS behaviors. 
Our evaluation further shows that improving coverage of difficult-to-detect attacks must be considered together with false positives on benign traffic, highlighting the need to evaluate both attack detection and false alarms when designing practical FBS defenses.

\section*{Acknowledgments}
We thank the reviewers and shepherd for their valuable comments. We would also like to thank Allison Lu, Aviva Smith, Bernardo B. P. Medeiros, Jordan Greene, Magdalena Pasternak, and Yungwoo Ko for their hospitality during the visit to the University of Florida. This work was supported by Institute of Information \& communications Technology Planning \& Evaluation (IITP) grant funded by the Korea government (MSIT) (No.RS-2024-00441762, Global Advanced Cybersecurity Human Resources Development).
  
  \bibliographystyle{plain}
  \bibliography{ref}
  \appendices
\section{Systematic Analysis of FBS Detection}
\label{appendix:detection}

\newcommand{\twoline}[2]{\begin{tabular}[c]{@{}c@{}}#1\\#2\end{tabular}}
\begin{table*}[t]
\centering
\begin{threeparttable}
\caption{Detection primitives used across FBS detection studies and systems.}
\label{tab:summary_of_fbs_detection}
\scriptsize

\setlength{\tabcolsep}{3pt} 
\begin{tabular}{cccccccccccccc}
\hline
\multirow{4}{*}{\hfill \twoline{Detection}{Technique}} & \multirow{4}{*}{\hfill Previous studies} & \multicolumn{10}{c}{Detection primitives} & \multirow{4}{*}{\hfill \twoline{Detection}{Timing}} & \multirow{4}{*}{\hfill \twoline{Detection}{entity}}\\ \cline{3-12} 
                                  &                                     &  \multicolumn{2}{c|}{Access} & \multicolumn{6}{|c|}{Protocol} & \multicolumn{2}{|c}{Physical} & & \\ \cline{3-12}
& & 
\begin{tabular}[c]{@{}c@{}}Signal\\strength\end{tabular} & 
\begin{tabular}[c]{@{}c@{}}RRC\\failure\end{tabular} & 
\begin{tabular}[c]{@{}c@{}}IMSI-exp.\\msgs.\end{tabular} & 
\begin{tabular}[c]{@{}c@{}}Reject\\msgs.\end{tabular} & 
\begin{tabular}[c]{@{}c@{}}Cell\\info.\end{tabular} & 
\begin{tabular}[c]{@{}c@{}}Re-\\direction\end{tabular} & 
\begin{tabular}[c]{@{}c@{}}Null\\cipher\end{tabular} & 
\begin{tabular}[c]{@{}c@{}}TA\\cmd.\end{tabular} & 
\begin{tabular}[c]{@{}c@{}}RF\\char.\end{tabular}& 
\begin{tabular}[c]{@{}c@{}}Timing\\error\end{tabular} & & \\
\hline
\multirow{18}{*}{\makecell[c]{Anomaly \\ Detection}} 
& Crocodile Hunter~\cite{crocodilehunter} & \textopenbullet &  &  &  & \textopenbullet &  &  &  &  &  & pre-hijack & \multirow{6}{*}{3rd party} \\ \cline{2-13}
& FBS-Radar$^\dagger$~\cite{li2017fbs} & \textopenbullet &  &  &  & \textopenbullet &  &  &  &  &  & post-attack &  \\ \cline{2-13}
& GSMK Overwatch~\cite{projectoverwatch} & &  &  &  & \textopenbullet & \textopenbullet & \textopenbullet &  &  &  & pre-hijack &  \\ \cline{2-13}
& IMSI-Catcher-Catcher$^\ddagger$\cite{dabrowski2014imsi} & \textopenbullet &  &  &  & \textopenbullet &  & \textopenbullet &  &  &  & mid-attack &  \\ \cline{2-13}
& Marlin~\cite{tucker2025detecting} &  &  & \textopenbullet &  &  &  &  &  &  &  & post-attack &  \\ \cline{2-13}
& SeaGlass~\cite{ney2017seaglass} & \textopenbullet &  & \textopenbullet &  & \textopenbullet & \textopenbullet & \textopenbullet &  &  &  & pre-hijack &  \\ \cline{2-14}
& Heijligenberg \etal~\cite{heijligenberg2024attacks} & \textopenbullet &  &  &  & \textopenbullet &  &  &  &  &  & post-attack & \multirow{6}{*}{Network} \\ \cline{2-13}
& Kara{\c{c}}ay \etal~\cite{karaccay2021network} & \textopenbullet &  &  &  &  &  &  &  &  &  & post-attack &  \\ \cline{2-13}
& Murat~\cite{nakarmi2021murat} & \textopenbullet &  &  &  & \textopenbullet &  &  &  &  &  & post-attack &  \\ \cline{2-13}
& Nakarmi \etal~\cite{nakarmi2022applying} & \textopenbullet &  &  &  &  &  &  &  &  &  & post-attack &  \\ \cline{2-13}
& Park~\cite{park2023we} & \textopenbullet & \textopenbullet &  &  & \textopenbullet &  &  &  &  &  & post-attack &  \\ \cline{2-13}
& Steig \etal~\cite{steig2016network} &  &  &  &  & \textopenbullet &  &  &  &  &  & post-attack &  \\ \cline{2-14}
& Apple's patent~\cite{apple_patent} & \textopenbullet &  &  &  & \textopenbullet & \textopenbullet &  & \textopenbullet & \textopenbullet $^\ast$ & \textopenbullet & post-hijack & \multirow{6}{*}{UE} \\ \cline{2-13}
& CellGuard~\cite{arnold2024catch} & \textopenbullet &  &  & \textopenbullet & \textopenbullet &  &  &  &  &  & mid-attack &  \\ \cline{2-13}
& EAGLE Security~\cite{eaglesecurity} & \textopenbullet &  &  &  & \textopenbullet &  &  &  &  &  & post-hijack &  \\ \cline{2-13}
& Huang \etal~\cite{huang2018identifying} & \textopenbullet &  &  &  &  &  &  &  &  &  & mid-hijack &  \\ \cline{2-13}
& LeopardSeal~\cite{peeters2023leopardseal} &  &  &  &  &  & &  & & \textopenbullet$^\sharp$ & & mid-attack &  \\ \cline{2-13}
& Rayhunter$^\flat$~\cite{rayhunter} &  &  & \textopenbullet &  & \textopenbullet & \textopenbullet & \textopenbullet &  &  &  & mid-attack &  \\ \hline
\multirow{4}{*}{\makecell[c]{Signature-based \\ Detection}} 
& FBSleuth~\cite{zhuang2018fbsleuth} &  &  &  &  &  &  &  &  & \textopenbullet &  & pre-hijack & 3rd party \\ \cline{2-14}
& Ali \etal~\cite{ali2019enabling} &  &  &  &  &  &  &  &  & \textopenbullet &  & mid-hijack & \multirow{3}{*}{UE} \\ \cline{2-13}
& FBSDetector$^\natural$~\cite{samin2024gotta} & \textopenbullet & \textopenbullet & \textopenbullet & \textopenbullet & \textopenbullet &  & \textopenbullet &  &  &  & post-attack &  \\ \cline{2-13}
& PHOENIX~\cite{echeverria2021phoenix} &  & \textopenbullet & \textopenbullet & \textopenbullet &  &  & \textopenbullet &  &  &  & post-attack &  \\ \hline

\end{tabular}

\begin{tablenotes}    
    \scriptsize
    \item Note that we do not cover tools that have not been maintained in the past decade, such as AIMSICD~\cite{aimsicd}, SnoopSnitch~\cite{snoopsnitch}, Darshak~\cite{darshak}, Cell Spy Catcher~\cite{spycatcher}, SITCH~\cite{wilson2016sitch}, and GSM Spy Finder~\cite{gsmspyfinder}, as they are no longer updated and their technical approaches have been largely incorporated into or superseded by more recent detection systems.
    \item[$\ast$] Monitoring Doppler effect for detecting mobile FBSs.
    \item[$\dagger$] Operating as a UE-assisted and cloud server-based system.
    \item[$\ddagger$] IMSI-Catcher-Catcher can be built on mobile UEs (M) and stationary 3rd party devices (S) to form a sensor network.
    \item[$\sharp$] Monitoring round trip timing (latency) of voice signals for detecting call interception by FBS.
    \item[$\flat$] Rayhunter can be built on Orbic mobile hotspots~\cite{orbic} and Qualcomm-based devices, equipped with a SIM card.
    \item[$\natural$] FBSDetector can operate either as an Android application on the UE or as standalone machine-learning models that process packet captures.
\end{tablenotes}

\label{tab:fbs_detection_primitives_full}

\end{threeparttable}
\end{table*}

In this section, we systematically investigate existing FBS detection mechanisms to establish the analytical foundation (\ie, indexing lens) for \devil's operational variants in~\autoref{ss:evasion}. We decompose detection techniques into underlying primitives -- observable features that reveal FBS presence. 

\autoref{tab:fbs_detection_primitives_full}, the full table from which ~\autoref{tab:fbs_detection_primitives} is derived, presents a comprehensive analysis of 22 existing FBS detection works. We focus on characterizing detection primitives across operational layers (\autoref{ss:primitives}); seven operational variants (V1-V7) in~\autoref{ss:evasion} are derived from these primitives combined with protocol specifications. We then examine how existing systems combine these primitives (\autoref{ss:hybrid}).

\subsection{Detection Primitives}
\label{ss:primitives}

We identify three domains where FBS behavior can be detected -- access-layer, protocol-layer, and physical-layer -- based on our unified model and the characteristics of cellular communications.

\nibf{Access-layer.}
This domain corresponds to the connection hijacking phase where FBS operations transition from passive observation and signal broadcasting to actively interacting with UEs to set up the link. In this process, the FBS disconnects UEs from legitimate cells and redirects them to itself, generating observable artifacts such as radio link failures.

\nibf{Protocol-layer.}
FBSs execute attacks over hijacked connections, a process that primarily aligns with the application phase and requires communication with UEs through standard cellular protocols. Such communication produces detectable patterns in signaling messages, reject sequences, and protocol violations. The standardized nature of cellular protocols thus offers a rich foundation for detection mechanisms based on message analysis and behavioral patterns.

\nibf{Physical-layer.}
Beyond the active interaction domains, physical-layer characteristics also provide useful indicators for FBS detection.
When the hardware setup of FBSs relies on relatively low-cost components with limited capabilities compared to commercial infrastructure, they exhibit distinctive RF characteristics and signal quality variations~\cite{zhuang2018fbsleuth}.

\subsubsection{Access-layer Primitives} 
\hfill \break
\nibf{Signal strength.}
A core behavior of FBSs during connection hijacking is attracting UEs by transmitting stronger signals than legitimate base stations~\cite{eff_surveillance}. This effect is particularly pronounced when hijacking is assisted by jamming: by suppressing legitimate cells, an FBS can more readily dominate UE measurements and induce selection via the capture effect~\cite{yang2019hiding,oh2024enabling,erni2022adaptover}. Accordingly, prior work has attempted to detect hijacking by monitoring whether received signal power exceeds an absolute threshold~\cite{arnold2024catch, li2017fbs} or by using the disappearance of neighboring legitimate cells (\eg, under jamming) as evidence of FBS activity~\cite{eaglesecurity}.

In addition to thresholding, some systems treat signal strength as a consistency/localization signal by (i) comparing reported Reference Signal Received Power (RSRP) or Reference Signal Received Quality (RSRQ) against expected values from a signal-strength map or known base station locations/topology~\cite{heijligenberg2024attacks, huang2018identifying, nakarmi2022applying}, or (ii) using multi-point signal-strength patterns to localize a suspected FBS~\cite{karaccay2021network}.

\begin{figure*}[t]
    \captionsetup{skip=3pt}
    \centering

    \begin{minipage}{0.75\textwidth}
        \centering
        \includegraphics[width=\linewidth]{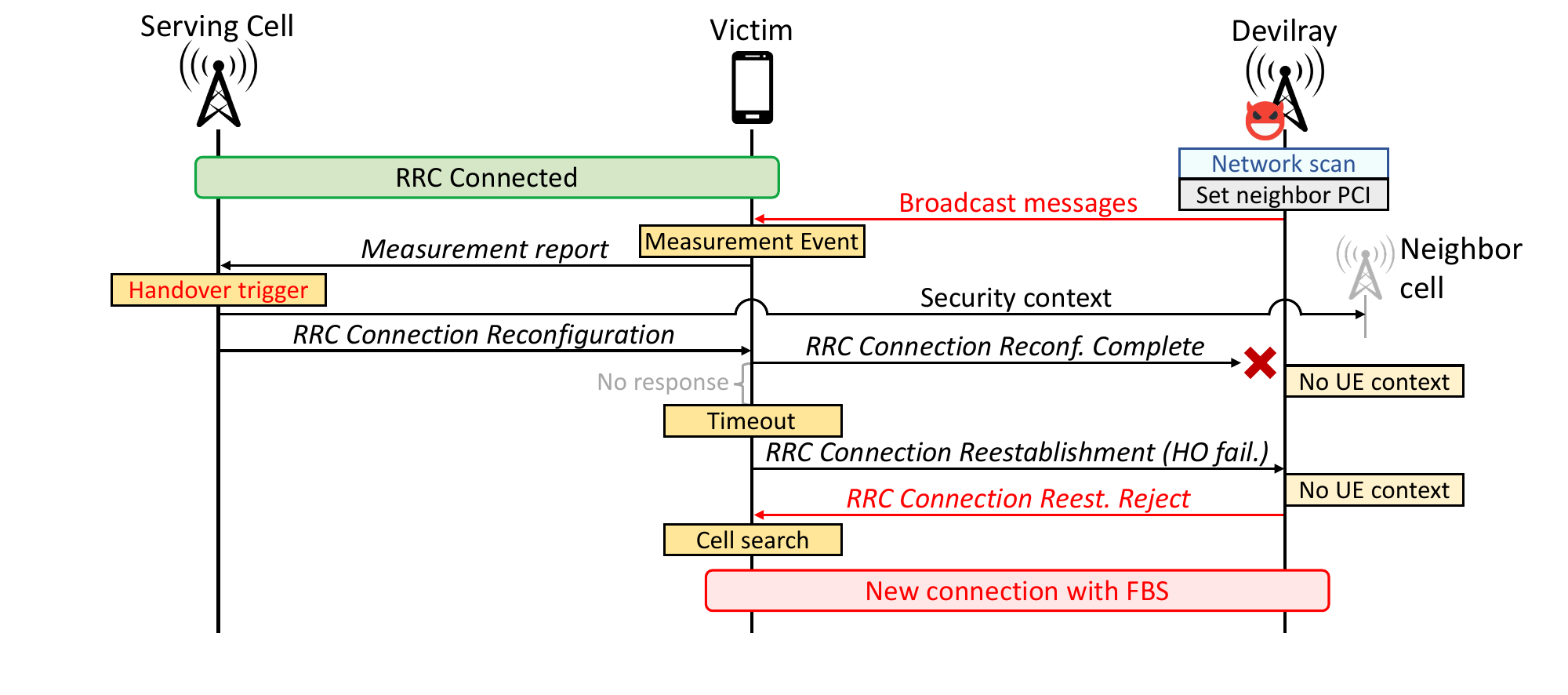}
        \caption*{\footnotesize (b) Handover }
    \end{minipage}

    \begin{minipage}{0.44\textwidth}
        \centering
        \includegraphics[width=\linewidth]{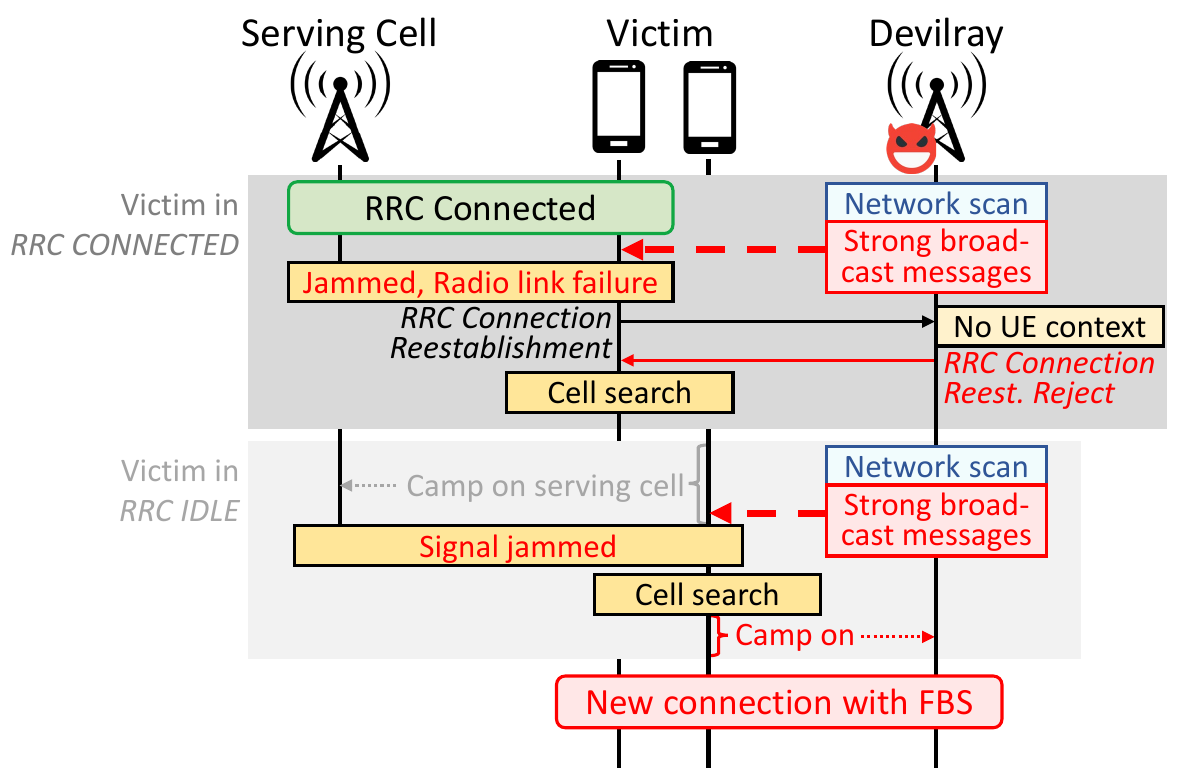}
        \caption*{\footnotesize (a) Jamming\phantom{-}}
    \end{minipage}
    \hfill
    \begin{minipage}{0.44\textwidth}
        \centering
        \includegraphics[width=\linewidth]{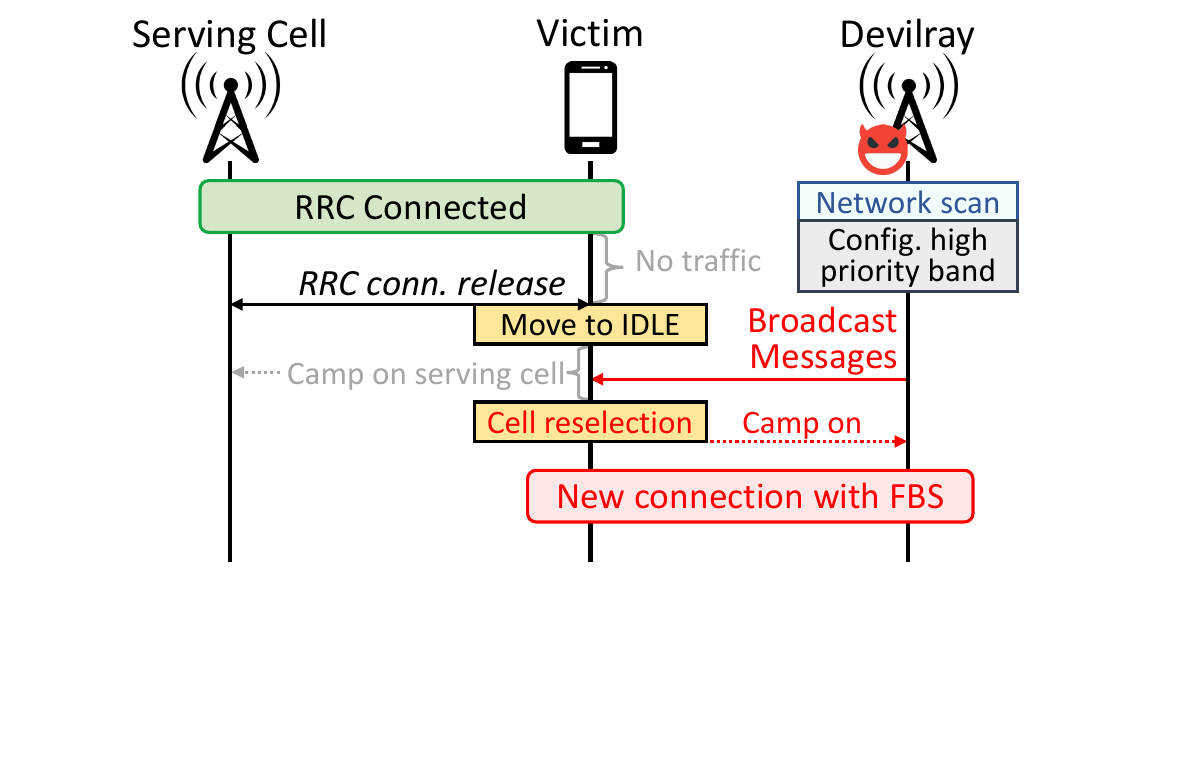}
        \caption*{\phantom{-}\footnotesize (c)  Cell reselection}
    \end{minipage}

    \vspace*{2mm}

    \caption{Message flow for three connection hijacking methods by FBSs}
    \label{fig:connection_hij_flow}
\end{figure*}

\nibf{RRC procedure failures.}
Victims experience RRC procedure failures during the connection hijacking phase that can reveal FBS presence. Specifically in jamming, according to our empirical study (\autoref{fig:connection_hij_flow}), the FBS disrupts the UE's link to the legitimate cell, causing Radio Link Failure (RLF) and leading to  \tts{RRC Connection Reestablishment Request} with ``other failure'' causes. In the case of handover, the UE sends  \tts{RRC Reconfiguration Complete} but the FBS cannot proceed due to missing context, triggering ``handover failure'' and subsequent  \tts{RRC Connection Reestablishment Request} with ``handover failure'' causes.

These failure patterns can help detect hijacking, though similar traffic patterns also arise in specific benign situations including UE in congestion or mobility. Prior works~\cite{echeverria2021phoenix,samin2024gotta,park2023we} focus on failure outcomes as standalone indicators, or treat them as part of a broader detector system.

\subsubsection{Protocol-layer Primitives} 
\hfill \break
\nibf{IMSI-exposing and reject messages.}
IMSI-exposing message patterns can reveal FBS activities aimed at identity tracking. Legitimate networks typically request IMSI sparingly and under specific conditions such as initial network attach or GUTI expiration~\cite{3gpp_24301}, while FBS generates excessive identity requests for IMSI-catching. To collect IMSIs, FBS can send various signaling messages to UEs, such as \tts{Identity Request}, \textit{NAS Reject}, and others~\cite{tucker2025detecting,3gpp_24301,3gpp_37340}.
 
Based on this knowledge, several prior works detect FBS activity by monitoring for IMSI-exposing messages on the control plane~\cite{tucker2025detecting,rayhunter,samin2024gotta,echeverria2021phoenix,ney2017seaglass}. Specifically, 
Marlin~\cite{tucker2025detecting} identifies suspicious situations of IMSI-catcher presence by monitoring statistical increases in 53 IMSI-exposing messages, including \textit{NAS Reject} with specific causes and \tts{Identity Request} messages.
Rayhunter~\cite{rayhunter} monitors IMSI-relevant messages including IMSI paging and IMSI \tts{Identity Request} messages.

Reject messages are another indicator of FBS activities, typically issued when authentication is required but cannot be completed, as FBSs lack the capability to correctly perform authentication procedures. This process disconnects UE links and returns them to legitimate networks after performing attacks.
As an attack purpose, they are also exploited for service denial and downgrade attacks along with IMSI-catching.
These patterns can be used to identify FBS activities by monitoring the frequency, cause values, and contextual abnormalities of reject messages~\cite{samin2024gotta,echeverria2021phoenix,arnold2024catch}.

\nibf{Cell information.}
FBS deployments can be detected by anomalies of several configurations. These include misconfigurations of cell information parameters, such as invalid TAC, cell IDs, or use of non-standard frequency bands that can be indicators of FBS. Monitoring broadcast messages can also be helpful, as FBSs may transmit malicious messages to influence UE behavior. For instance, cell reselection priorities or barred cell status in SIB messages can mislead UEs into connecting to an FBS or prevent them from connecting to legitimate networks. Additionally, malicious paging messages may be used to elicit UE responses, enabling identity exposure or DoS through repeated, targeted paging.

Based on these observations, detectors flag an FBS when the observed cell information appears abnormal or distinguishable from legitimate cells. Common examples include cells advertising PLMN/TAC/cell IDs that are different from neighbors or not registered in a database~\cite{arnold2024catch,eaglesecurity,crocodilehunter,li2017fbs,heijligenberg2024attacks,dabrowski2014imsi,nakarmi2021murat,rayhunter}, as well as cells whose broadcast information is inconsistent such as containing unusually small timer values or missing neighbor-cell information~\cite{apple_patent,ney2017seaglass, projectoverwatch}.

\nibf{3G/2G redirection and null ciphers.}
Detection systems can also monitor redirection messages or null cipher usage~\cite{samin2024gotta,echeverria2021phoenix,rayhunter,apple_patent}. These indicators target specific application scenarios: redirection to legacy networks and SMS injection or eavesdropping through encryption downgrade.

Redirection occurs through multiple mechanisms specified in 3GPP standards. Regarding RRC connection management, \tts{RRC Connection Release} messages may contain \tts{redirectedCarrierInfo} to force cell selection steering UEs toward specific RATs or frequencies.
Additionally, SIB messages can specify legacy technology priorities to induce inter-RAT reselection.

On the one hand, null cipher can be exploited to eavesdrop on UE communications. During the RRC connection and NAS Attach procedure, an adversary can propose null encryption algorithms via \tts{Security Mode Command} to disable encryption and expose UE data.

Note that redirection and ciphering changes can occur in benign operations. For example, a UE autonomously performs cell search on lower-generation RATs if it fails to find any suitable cells~\cite{3gpp_36304}. Additionally, networks can use redirection for congestion management and load balancing, and ciphering changes for emergency calls.

\subsubsection{Physical-Layer Primitives}
\hfill \break
\nibf{RF characteristics.}
FBS signals might exhibit unique RF characteristics that can be used to differentiate them from legitimate network transmissions. These include signal bandwidth variations, frequency stability (center frequency offset), time synchronization offset, phase characteristics, and magnitude deviations in pilot signals, reflecting hardware specifications and transmission conditions. For example, FBSleuth~\cite{zhuang2018fbsleuth} demonstrates this detection approach by measuring these physical-layer features. Other approaches~\cite{peeters2023leopardseal,ali2019enabling} utilized signal noise distributions or round-trip time analysis for additional RF characteristics.

\nibf{Transmission timing error.}
Apple's patent~\cite{apple_patent} describes using clock synchronization capabilities and Doppler effects for FBS detection. Building on this concept, transmission timing error can serve as a detection primitive. Legitimate cells typically achieve precise clock synchronization through GPS or external clock sources from their mobile operators, which provide accurate timing information to all cells in the network and ensure coordinated operation across the network. This synchronization is critical for maintaining network stability, enabling seamless handovers between cells, compensating for timing/frequency offsets, and preventing inter-cell interference~\cite{3gpp_36401,3gpp_38401}.

In contrast, resource-constrained FBS relies on less accurate internal clocks, which introduce timing misalignment and signal distortions such as frequency offset. By focusing on the clock synchronization capabilities of legitimate cells, the inherent timing misalignment in FBS transmissions can be utilized to differentiate them from legitimate cells. 

\subsection{Existing Detection Techniques}
\label{ss:hybrid}
We now examine how existing detection systems combine these primitives in practice.
We analyze existing detection systems across three dimensions:

\nibf{Detection technique}. Anomaly detection identifies deviations from normal network behavior while signature-based detection matches observed patterns against known attack signatures.

\nibf{Detection timing.}
We define five phases: pre-, mid-, and post-hijacking (before any FBS interaction, during redirection, and after UE connection but before attacks); and mid-/post-attack (during misuse and after-the-fact detection). Early phases allow prevention; later ones support confirmation and forensics.

\nibf{Detection entity.}
FBS detection can occur on the UE side, network side, or via third-party systems. UE-side detection runs on user devices, often via apps, and relies on local observation. Network-side detection uses MNO infrastructure to aggregate data from UEs or base stations. Third-party detection leverages external systems -- independent of both MNO and UE -- using dedicated hardware or data from multiple sources.

We organize our analysis primarily by detection technique as this captures the core operational logic underlying different systems.

\subsubsection{Anomaly Detection}
\hfill \break
Identifying deviations from normal behavior in legitimate networks, suspicious patterns can be detected and flagged as potential FBS activity. This approach typically focuses on monitoring emerged cells showing abnormal signal strength, message transmissions, unexpected cell location and unusual cell information.

CellGuard~\cite{arnold2024catch} determines the presence of FBS by scoring the likelihood of a cell being malicious based on the Apple Location Service (ALS) database including cell information. With cell location information in the ALS database, CellGuard running on iPhone verifies the existence of connected cells and cell configurations (frequency channel and cell ID) against the database. Additionally, it calculates the distance between the UE and approximate location of the cell and monitors network reject messages, bandwidth, and signal power level to determine anomalies in cell behavior.
Crocodile Hunter~\cite{crocodilehunter}, developed by the Electronic Frontier Foundation (EFF), also scores suspiciousness of cells by collecting their configurations (PLMN, TAC, cell ID, frequency channel) and its location to compare with the open cell database, OpenCelliD~\cite{opencellid}, and monitoring signal power fluctuations.

EFF recently released Rayhunter~\cite{rayhunter}, a new FBS detection tool that monitors specific indicators including IMSI paging, IMSI \tts{Identity Request}, SIB and \tts{RRC Connection Release} messages associated with 2G redirection, and null ciphers over \tts{Security Mode Command} and \tts{RRC Connection Reconfiguration} messages to identify FBS activities. Marlin~\cite{tucker2025detecting} identifies IMSI-Catchers by monitoring the statistical increase of 53 IMSI-related messages, including \textit{NAS Reject} with specific causes, and \tts{Identity Request} messages.

EAGLE Security~\cite{eaglesecurity}, developed by Int64 Team, also operates as a UE-side Android application that validates cell configurations against the OpenCelliD database similar to Crocodile Hunter. The app evaluates multiple criteria to assign ``wiring probability'' scores: database presence verification, distance comparison between recorded and measured cell locations, PLMN validity checks, and analysis of whether multiple cells or only single cells are visible at the current location. Violations of these criteria increase the suspicion score for potential FBS detection.

FBS-Radar~\cite{li2017fbs} collects recently connected cell information (\eg, cell ID, signal strength, and connection time), MAC addresses of nearby WiFi access points, and SMS contents from UEs and stores this data in a cloud server. 
Based on the collected data, FBS-Radar analyzes cell information using a location database of cells and WiFi, and examines SMS contents using machine learning models to identify FBS activities sending spam SMSes.

Heijligenberg \etal~\cite{heijligenberg2024attacks} suggested an FBS detection method monitoring Tracking Area Update (TAU) messages with dummy TAC values and comparing cell power from UE measurement reports with the signal strength map at the network side. Similarly, Huang~\etal~\cite{huang2018identifying} and Nakarmi~\etal~\cite{nakarmi2022applying} proposed methods to identify FBS by modeling expected signal strength received at the UE. Additionally, Kara{\c{c}}ay \etal~\cite{karaccay2021network} introduced an FBS localization method using observed signal strengths of cells in UE measurement reports and trilateration with a signal propagation model.
Other approaches~\cite{steig2016network,nakarmi2021murat,park2023we} collect measurement reports from UEs, including neighboring cell lists, signal strength, or RLF reports, and analyze them against network databases to identify FBS activities.

IMSI-Catcher-Catchers~\cite{dabrowski2014imsi}, Overwatch~\cite{projectoverwatch} and SeaGlass~\cite{ney2017seaglass} are designed to collect cell information or monitor IMSI-exposing messages through scalable deployments of UEs or sensor networks and build a database of regional cells to identify suspicious activities.
These systems rely on the assumption that legitimate base stations exhibit consistent and predictable behaviors over time and across geographic locations. They collect information such as cell IDs, signal strength, broadcast parameters, and encryption status, along with precise GPS coordinates. By aggregating this data, they construct a baseline model of known legitimate cells. Detection is then performed by identifying inconsistencies such as previously unseen cell IDs and unexpected changes in broadcast configurations and cell locations when compared to the reference database.

As a comprehensive documentation, Apple's patent~\cite{apple_patent} presents a detection technique that combines various detection primitives, including monitoring signal strength, cell information, TA commands, time synchronization capability, and Doppler effect for detecting mobile FBSs. However, the document only theoretically describes this detection technique without providing experimental results or practical implementation details. Separately, LeopardSeal~\cite{peeters2023leopardseal} introduces a method to detect call interception by FBS by monitoring the round trip timing of voice signals.

\subsubsection{Signature-based Detection}
\hfill \break
As another methodology, attack signatures can be used to identify FBS activities. These signatures are typically based on known attack patterns, such as specific message sequences or behaviors associated with FBS operations. By comparing observed behaviors against these signatures, detection systems can identify FBS attacks.

PHOENIX~\cite{echeverria2021phoenix} utilizes a signature-based approach to detect FBS attacks by monitoring specific message patterns, including RLF reports. It focuses on identifying attack signatures, which consist of undesired message sequences not present in benign cellular behaviors. FBSDetector~\cite{samin2024gotta} employs signatures of abnormal activities and multi-step attacks of FBS with large-scale datasets. Based on machine learning techniques, it is also designed to cover unseen and reshaped FBS attacks.

FBSleuth~\cite{zhuang2018fbsleuth} is a representative work that employs RF fingerprinting as signatures to distinguish signals between FBS and legitimate cells in 2G networks. Based on the hardware imperfections inherent in FBS equipment, it models fingerprints with RF characteristics such as modulation errors and instantaneous phase and frequency, and applies machine learning techniques to classify these fingerprints.  Ali~\etal~\cite{ali2019enabling} proposed a similar approach to detect FBS by analyzing signal distortion derived from hardware limitations of FBS. 

\section{Dependency Checker}

\section{Derivation of \devil Instances}
\label{appendix:numbers}
The total number of unique instances ($N_{\text{total}}$) is derived as follows:
$$
N_{\text{total}} = N_{\text{launch}} \times N_{\text{hijack}} \times N_{\text{app}} = 32 \times 3 \times 27 = 2,592
$$
\begin{itemize}
    \item \textbf{$N_{\text{launch}}$ (32)}: $2$ (broadcast parameter adaptation) $\times$ $2$ (cell iteration) $\times$ $2$ (paging reproduction) $\times$ $2$ (TA diversification) $\times$ $2$ (hardware compensation). Each factor represents a binary configuration for its respective strategy (\eg, two broadcast parameter adaptation: unadapted or adapted). 
    \item \textbf{$N_{\text{hijack}}$ (3)}: Jamming, cell reselection, and handover.
    \item \textbf{$N_{\text{app}}$ (27)}: $3$ (targeting) $\times$ $9$ (app variation). App variations are derived from $3$ (IMSI-catching) $+$ $2$ (location tracking) $+$ $1$ (DoS) $+$ $3$ (Redirection and SMS injection). 
\end{itemize}
This counting is conservative: we enumerate only high-level factors that admit a clean Cartesian-product structure (32$\times$3$\times$27), and intentionally coarsen other within-variant degrees of freedom instead of expanding them.
For example, while some variants admit multiple standard-defined choices (\eg, different reject causes), we treat them as a single category to keep the space interpretable.

\section{Connection Hijacking and Applications}
\label{appendix:eval_efficacy}

In this section, we demonstrate \devil's effectiveness on connection hijacking and applications.
Our setup consists of a laptop running \devil software connected to two USRP B210s. Target devices include seven COTS UEs -- Galaxy S10e, S20, S21, S23, iPhone XS, 12 and 13 Pro -- each equipped with commercial SIM cards issued for research use. To ensure generalizability and reproducibility, we perform these experiments across two countries on different continents.

\nibf{Connection hijacking.}
We evaluate the effectiveness of \devil's connection hijacking mechanisms on COTS UEs to confirm their practical impact and to characterize their operational requirements. 
Target UEs are controlled into idle or connected states according to the tested mechanism: cell reselection trials start with UEs camping in idle mode, handover trials are performed during active voice sessions to maintain connected mode, and jamming is tested against both idle and connected UEs. Each experiment is repeated ten times across five relative power margins, up to 30 dB.

\autoref{tab:devil_conn_hjk_eval} presents the connection hijacking success rates for different relative signal power levels across all methods. The table show the results against one representative device, the Galaxy S23. While jamming-based hijacking requires a minimum of 20\,dB signal advantage and achieves a 100\% success rate at 30\,dB, cell reselection and handover-based hijacking require much less relative power. Cell reselection is effective with as little as 5\,dB advantage, while handover to \devil is performed even with just a 3\,dB stronger signal.

\autoref{tab:fbs_hij_requirement} summarizes the operational requirements for each hijacking mechanism.
Across all seven tested UEs, \devil succeeds in performing connection hijacking under the expected conditions, though the success rate for each mechanism varies depending on the relative power margin.
While jamming-based hijacking requires a minimum of 20\,dB signal advantage, cell reselection and handover-based hijacking require much less relative power. Cell reselection is effective with as little as 5\,dB advantage, while handover to \devil is performed even with just a 3\,dB stronger signal.
Jamming influences both idle and connected UEs on the serving frequency, requiring a high relative power advantage. Handover applies only to connected UEs and can succeed with a subtle power margin while reusing neighbor PCIs. Cell reselection targets only idle UEs and leverages higher-priority frequency settings with a modest power margin.

\begin{table}[t]
\setlength{\tabcolsep}{6pt}
    \caption{\devil's connection hijacking success rate (\%)}
    \label{tab:devil_conn_hjk_eval}
    \centering
    \resizebox{\linewidth}{!}{
    \scriptsize
    \begin{tabular}{lccccc}
    \hline
    \textbf{Relative power} & \textbf{3\,dB} & \textbf{5\,dB} & \textbf{10\,dB} & \textbf{20\,dB} & \textbf{30\,dB} \\
    \hline
    Jamming & 0 & 0 & 0 & 30 & 100 \\
    Handover & 20 & 70 & 100 & 100 & 100 \\
    Cell reselection & 0 & 20 & 70 & 100 & 100 \\
    \hline
    \end{tabular}
    }
    
    \begin{tablenotes}
    \begin{minipage}{.95\columnwidth}
        \scriptsize
        \item[] * Relative power indicates the signal strength advantage compared to legitimate serving cells of target UEs (shown here for representative UE, Galaxy S23).
        \item[] * We conducted 10 trials for each method and relative power.
        \end{minipage}
    \end{tablenotes}
\end{table}

\begin{table}[t]
    \centering
    \caption{Connection hijacking requirements}
    \label{tab:fbs_hij_requirement}
    \resizebox{\linewidth}{!}{
    \begin{threeparttable}
    
    \begin{tabular}{c|c|c|c|c}
        \hline
        Requirements & \textbf{Jamming} & \textbf{Handover} & \multicolumn{2}{|c}{\textbf{Cell reselection}}     \\ \hline
    Frequency                    & Serving & Any & Serving$^\ast$ & Higher-priority  \\ \hline
    PCI                     & Any & Neighbor & Different$^\ast$ & Any \\ \hline
    Relative power$^\dagger$     & $>$ 20\,dB & $>$ 3\,dB$^\ddagger$ & \multicolumn{2}{c}{$>$ 5\,dB$^\ddagger$}  \\ \hline
    UE state      & Any state   & Connected  & \multicolumn{2}{c}{Idle}  \\
    \hline
    \end{tabular}%
    \begin{tablenotes}
    \scriptsize
    \item[] Frequency and PCI rows describe whether Devilray should match or differ from the UE’s serving cell – \ie, the cell to which the UE is currently connected or camping on – in terms of operating frequency and PCI.
    \item[] $\ast$ For the idle UE camping on the serving cell with the highest priority frequency, Devilray's PCI must differ from the serving cell's, while operating on the same highest priority band to be perceived as a distinct, higher-power cell for reselection exploitation. 
    \item[] $\dagger$ See \autoref{tab:devil_conn_hjk_eval} for detailed results.
    \item[] $\ddagger$ Depending on the network configuration~\cite{3gpp_36331,3gpp_36304}.
    \end{tablenotes}
\end{threeparttable}
}
\end{table}

\nibf{Application.}
We evaluate on-board attacks of \devil (\ie, IMSI-catching, location tracking, 2G redirection, SMS injection, and DoS) with a jamming as hijacking. 

As a result, all IMSI-catching modes of \devil successfully extract IMSIs from target UEs immediately after connection hijacking. We further confirm that it can continuously trigger UE traffic with IMSI paging and \tts{Identity Request}, enabling location estimation through measurements of arrival signal strength. 
For 2G downgrades and SMS injection, five COTS UEs (Galaxy S10e, S21, iPhone XS, 12 Pro, and 13 Pro) were successfully redirected to \devil's 2G cell and received malicious SMS messages over the unencrypted link. The remaining two devices were redirected to the 2G band but did not connect to \devil's 2G cell.

For the DoS attack, we confirm that every UE that receives \tts{Attach Reject} with cause \#22, is unable to reconnect to the mobile network for up to 30 minutes. Notably, we observe that the disruption is not resolved by common user actions such as toggling airplane mode or rebooting the device. Furthermore, after recovery, reapplying the same attack re-induces the disruption, confirming its repeatability. We execute up to 5 consecutive DoS attempts, all of which are successful. 

\section{Measurements of RF Characteristics}
\label{appendix:rf}
\setlength{\tabcolsep}{3pt}

\begin{table}[h]
\centering
\caption{Measurement results of RF characteristics}
\label{tab:rf_fingerprint}
\begin{threeparttable}

\resizebox{.95\columnwidth}{!}{%
\begin{tabular}{lcccc}
\hline
\textbf{Target} & \textbf{Condition} & \textbf{CFO (Hz)} & \textbf{Sync. error (ns)} & \textbf{Mag. error (\%)} \\
\hline
\multirow{2}{*}{MNO1}
  & Day1  & \num{-1416} $\pm$ \num{154} & \num{1493}  $\pm$ \num{77}  & \num{1.3} $\pm$ \num{0.3} \\
  & Day7  & \num{-1402} $\pm$ \num{119} & \num{1129}  $\pm$ \num{91}  & \num{1.2} $\pm$ \num{0.1} \\
\hline
\multirow{2}{*}{MNO2}
  & Day1  & \num{-1316} $\pm$ \num{115}  & \num{1927} $\pm$ \num{294} & \num{1.7} $\pm$ \num{0.1} \\
  & Day7  & \num{-1419} $\pm$ \num{201}  & \num{2526} $\pm$ \num{889} & \num{1.3} $\pm$ \num{0.3} \\
\hline
\multirow{2}{*}{MNO3}
  & Day1  & \num{-1541} $\pm$ \num{101}  & \num{238}  $\pm$ \num{81}  & \num{1.9}  $\pm$ \num{0.5} \\
  & Day7  & \num{-1501} $\pm$ \num{122}  & \num{382}  $\pm$ \num{63}  & \num{2.7}  $\pm$ \num{0.2} \\
\hline
\multirow{3}{*}{\devil$^\ast$}       
& B210 & \num{-475}  $\pm$ \num{145}  & \num{-866} $\pm$ \num{59}  & \num{16.9} $\pm$ \num{2.9} \\
& X310 & \num{-4130}  $\pm$ \num{93}  & \num{1937} $\pm$ \num{59}  & \num{2.1} $\pm$ \num{0.2} \\
& RF manip. & \num{-1346}  $\pm$ \num{176}  & \num{1563} $\pm$ \num{492}  & \num{3.3} $\pm$ \num{0.4} \\
\hline
Amari Callbox$^\ast$ & - & \num{-3376} $\pm$ \num{377}  & \num{8147} $\pm$ \num{622} & \num{6.4}  $\pm$ \num{0.9} \\
\hline
C-FBS$^\ast$ & - & \num{-1443} $\pm$ \num{116} & \num{506}  $\pm$ \num{12}  & \num{12.1} $\pm$ \num{1.6} \\
\hline
\end{tabular}
}%
\begin{tablenotes}
    \scriptsize
    \begin{minipage}{.95\columnwidth}
    \item[] $\ast$ Measurements on Devilray, Amari Callbox, and C-FBS are conducted in the controlled room with conductive mesh shielding for ethical considerations. In contrast, all passive measurements on MNOs are performed in normal conditions.
    \end{minipage}
\end{tablenotes}
\end{threeparttable}
\end{table}

FBSleuth~\cite{zhuang2018fbsleuth} introduces RF fingerprinting for 2G FBS detection, proposing to fingerprint FBSs based on signal distortions stemming from hardware imperfections. In this paper, we extend this primitive to LTE FBS detection and evaluate it under hardware compensation (V7).

We utilize three features as RF characteristics, selected based on FBSleuth~\cite{zhuang2018fbsleuth} and channel estimation methods in srsRAN~\cite{srsran}: CFO, time synchronization error, and magnitude error. CFO captures frequency mismatches between the received PSS (Primary Synchronization Signal) and center frequency, while time synchronization error is the timing offset of received pilot signals, and magnitude error indicates amplitude deviations in pilot signals caused by modulation imperfections.

To examine RF characteristics' effectiveness in LTE, we measure three features from six operational cells across three MNOs, \devil, Amari Callbox~\cite{amari}, and the C-FBS. Measurements on MNOs are taken on Day~1 and Day~7 at the same location to capture temporal variations. Each measurement is 10 seconds long and repeated 10 times. We utilize LTESniffer's~\cite{hoang2023ltesniffer} built-in functions for CFO and synchronization error measurements, while extending it to measure magnitude error with FBSleuth's~\cite{zhuang2018fbsleuth} methodology. 
As a result, we confirm that the RF characteristics of the three MNOs remain consistent between Day 1 and Day 7. Additionally, \devil, Amari Callbox, and the C-FBS exhibit distinct patterns that differentiate them from legitimate cells (see \autoref{tab:rf_fingerprint} for detailed numbers).

For intuitive interpretation, we visualize the results using PCA. PCA is first performed using Day 1 and 7 measurements from legitimate cells, combined with measurements from \devil with B210, and Amari Callbox, to define the projection space. Subsequently, measurements from the C-FBS and \devil with X310 are projected to validate the projection space and evaluate the effectiveness of this detection primitive. \autoref{fig:pca_figure} shows that legitimate cells form tight clusters, while \devil, Amari Callbox, and the C-FBS each form separate, distinguishable clusters.

Then we project measurement results of \devil with RF tuning to explore the effect of hardware compensation (V7). Specifically, it applies software-based RF characteristic manipulation by deliberately introducing signal distortions, such as injecting CFO. The resulting measurements clustered together with those of legitimate cells, showing that RF tuning can influence RF characteristic-based detection.

\clearpage

\end{document}